\documentclass[useAMS,usenatbib,usegraphicx]{mn2e}
\usepackage{epsfig}
\usepackage{aas_macros} 
\usepackage{amsmath}
\usepackage{amssymb}
\usepackage{appendix}
\usepackage{booktabs}
\usepackage{upgreek}
\usepackage{url}
\usepackage[usenames]{color}
\newcommand{\ion}[2]{{#1}\,{\sc #2}}

\usepackage{graphicx}
\usepackage{rotating}
\usepackage{color}
\usepackage{pifont}
\usepackage{longtable}
\usepackage{pdflscape,lscape}
\usepackage{natbib}

\title[The $\delta$~Scuti visual binary HD\,21190]{An astrometric and spectroscopic study of the $\delta$\,Scuti variable HD\,21190 and its wide companion CPD\,$-$83$^{\circ}$\,64B}
\author[E.\,Niemczura et al.]{E.\,Niemczura$^1$\thanks{E-mail: niemczura@astro.uni.wroc.pl},
R.-D.~Scholz$^2$, S.~Hubrig$^2$, S.~P.~J\"arvinen$^2$, M.~Sch\"oller$^3$, I.~Ilyin$^2$,\and F. Kahraman Ali\c{c}avu\c{s}$^4$\\
$^1$ Instytut Astronomiczny, Uniwersytet Wroc{\l}awski, Kopernika 11, 51-622 Wroc{\l}aw, Poland\\
$^2$ Leibniz-Institut f\"ur Astrophysik Potsdam (AIP), An der Sternwarte~16, 14482~Potsdam, Germany\\
$^3$ European Southern Observatory, Karl-Schwarzschild-Str.~2, 85748~Garching, Germany\\
$^4$ Canakkale Onsekiz Mart University, Faculty of Sciences and Arts, Physics Department, 17100, Canakkale, Turkey
}

\begin{document}

\def\kms{{km\,s}$^{-1}$}
\def\ksi{$\xi_{\rm t}$}
\def\macro{$\zeta_{\rm RT}$}
\def\rad{$v_{\rm r}$}
\def\ebv{$E(B-V)$}

\def\teff{{T}_{\rm eff}}
\def\logg{\log g}
\def\micro{\xi_{\rm t}}
\def\vsini{v\sin i}

\date{Accepted ... Received ...; in original form ...}

\pagerange{\pageref{firstpage}--\pageref{lastpage}} \pubyear{2002}

\maketitle

\label{firstpage}

\begin{abstract}
Although pulsations of $\delta$~Scuti type are not expected among Ap stars from a theoretical point of view,
previous observations of the known $\delta$~Scuti star HD\,21190 indicated a spectral classification F2\,III SrEuSi:, making it the most evolved Ap star known.
Our atmospheric chemical analysis based on recent HARPS observations confirms the presence of chemical peculiarities in HD\,21190. 
This star is also the only target known to host a magnetic field along with its $\delta$~Scuti pulsation properties.
Using an astrometric analysis, we show that HD\,21190 forms a physical binary system with the companion CPD\,$-$83$^{\circ}$\,64B. 
The presented astrometric and spectroscopic study of the binary components is important to understand
the complex interplay between stellar pulsations, magnetic fields, and chemical composition.
\end{abstract}


\section{Introduction}

A previous analysis of time-series photometry and spectroscopic material of the {\it Hipparcos} variable star HD\,21190 by \citet{2001MNRAS.326..387K} indicated a spectral classification F2\,III~SrEuSi:, making this star the most evolved Ap star known to date. Moreover, they discovered that HD\,21190 shows pulsations of $\delta$~Scuti type, which are unusual among Ap stars. According to the authors, it was also one of the most evolved $\delta$~Scuti stars known. Ultraviolet and Visual Echelle Spectrograph (UVES) high time-resolution observations of HD\,21190 were used by \citet{2008CoSka..38..411G} and \citet{2008MNRAS.384.1140G} to search for pulsational line profile variations. The authors detected in several line profiles three peaks moving smoothly towards the red with a speed of $21$\,km\,s$^{-1}$. Such a behaviour is typical for non-axisymmetric non-radial pulsations. The combination of the cool Ap spectral type and the high-degree $\delta$~Scuti pulsation made HD\,21190 an important target for a more in depth study.

Noteworthy, pulsations of $\delta$~Scuti type among Ap stars are not expected from a theoretical point of view. The models of \citet{2005MNRAS.360.1022S} led to a clear prediction that lower radial overtone pulsation modes typical in $\delta$~Scuti stars are suppressed by the magnetic field in Ap stars. On the other hand, in the Hertzsprung-Russell (H-R) diagram, the coolest members of the Ap stars, the so-called rapidly oscillating Ap stars (roAp stars) with typical pulsation periods between $6$ and $24$\,min appear in the same instability strip as the $\delta$~Scuti stars. According to theoretical considerations, magnetic Ap stars should not be observed to pulsate simultaneously in low radial overtone $p$ modes characteristic for the $\delta$~Scuti stars and high overtone $p$ modes characteristic for the roAp stars. The typical pulsation periods of $\delta$~Scuti stars range from $20$\,min to $0.3$\,d and it is assumed that the $\kappa$~mechanism acting in the second ionization zone of helium is responsible for their pulsational behaviour. Since the pulsation periods of roAp stars and $\delta$~Scuti stars partially overlap, the combination of the cool Ap spectral type and high-degree $\delta$~Scuti pulsations makes HD\,21190 an excellent target for the study of the presence of a magnetic field in its atmosphere.

\citet{2008MNRAS.386.1750K} used the FOcal Reducer low dispersion Spectrograph (FORS\,1; \citealt{Appenzeller1998}) mounted on the $8$\,m Kueyen telescope of the VLT in spectropolarimetric mode with the GRISM 600B and a slit width of $0$\farcs$4$ to achieve a spectral resolution of about $2000$. The analysis of these data revealed the presence of a rather weak longitudinal magnetic field $\left<B_z\right>=47\pm13$\,G. \citet{2016IBVS.6174....1H} re-observed HD\,21190 with FORS\,2 in 2016 March, using the same instrumental setup as \citet{2008MNRAS.386.1750K}, and detected a magnetic field in HD\,21190 at a significance level of more than $4\sigma$ ($\left<B_{\rm z}\right>_{\rm all}= -254\pm59$\,G) using for the measurements the entire spectrum, and at a significance level of more than $3\sigma$ ($\left<B_{\rm z}\right>_{\rm hyd}= -237\pm75$\,G) using the hydrogen lines. This suggests that the magnetic field is variable with rotation, as is typical for Ap stars. However, further observations are necessary to determine the magnetic field geometry and the polar field strength of this star.

We note that there is no other known case of an Ap star with a detected magnetic field that shows $\delta$~Scuti pulsations. \citet{2015MNRAS.454L..86N} recently obtained two spectropolarimetric observations of the {\em Kepler} $\delta$~Sct -- $\gamma$~Dor hybrid candidate HD\,188774. One observation showed a weak signal of about $76$\,G. However, the target did not show any Ap characteristics: the abundance analysis did reveal neither the presence of chemical peculiarity nor of chemical spots. On the other hand, indications that a few Ap stars can show a hybrid $\delta$~Sct -- $\gamma$~Dor variability were reported by several authors, e.g.\ by \cite{2011MNRAS.410..517B} who used {\em Kepler} observations of stars in the $\delta$~Scuti instability strip.

The $\delta$~Scuti star HD\,21190 is so far the only target proven to host a magnetic field along with its $\delta$~Scuti pulsation properties. Up to now, only \cite{2001MNRAS.326..387K} carried out an abundance study for this object, using low-resolution spectra. Obviously, a high-resolution spectroscopic study is necessary to confirm the presence of chemical anomalies typical for cool Ap stars and to improve our understanding of the interplay between the magnetism and the physical processes taking place in the atmospheres of pulsating stars. 

In this work, our recent HARPS observations \citep{mayor2003} are used to characterize the abundances of $24$ chemical elements in the atmosphere of HD\,21190. In addition to observations of HD\,21190, we also obtained HARPS observations of the visual nearby companion CPD\,$-$83$^{\circ}$\,64B at an angular distance of about $19$\arcsec{}. The atmospheric parameters and chemical abundances of the companion were also determined. Additionally, the question whether HD\,21190 and CPD\,$-$83$^{\circ}$\,64B have a common proper motion and form a physical wide binary system has not been discussed in the past. In the following sections we present all available astrometric measurements obtained for both stars including those from {\it Hipparcos} and {\it Gaia} to investigate their proper motions and space velocities and discuss the results of our chemical abundance analysis obtained using high-resolution HARPS spectra.

\section{Astrometric measurements}
\label{sect:astro}

The two stars HD\,21190 and CPD\,$-$83$^{\circ}$\,64B are listed in the Washington Double Star (WDS) catalogue \citep{mason01} with a first satisfactory observation of their angular separation ($\sim15$\arcsec{}) in 1835. In the recently published catalogue of $\delta$~Scuti stars in binaries, \citet{liakos16} included CP\,Oct (HD\,21190) as a visual binary with an unspecified orbital period. In the following analysis of available astrometric measurements, we refer to HD\,21190 and CPD\,$-$83$^{\circ}$\,64B as A and B, respectively.

\subsection{Common proper motion} 
\label{SubSect_cpm}

\begin{table}
\caption{Proper motions of HD\,21190 (A) and CPD\,$-$83$^{\circ}$\,64B (B).} 
\label{table_pm}             
\centering                   
\begin{tabular}{cccl}       
\toprule
\multicolumn{1}{c}{Object} &
\multicolumn{1}{c}{$\mu_{\alpha}\cos{\delta}$ (error)} &
\multicolumn{1}{c}{$\mu_{\delta}$ (error)} &
\multicolumn{1}{c}{Source} \\
&
\multicolumn{1}{c}{[mas/year]} &
\multicolumn{1}{c}{[mas/year]} &
\\
\midrule
A & $-8.8$  (2.6) & $+21.5$  (1.0)  & ACT$^{a}$ \\
\smallskip                          
B & $-6.5$  (1.6) & $+16.4$  (5.2)  & ACT$^{a}$ \\
A & $-8.8$  (2.0) & $+19.2$  (1.8)  & TRC$^{a}$ \\
\smallskip                          
B & $-7.2$  (2.3) & $+14.9$  (3.6)  & TRC$^{a}$ \\
A & $-8.1$  (1.6) & $+19.4$  (1.6)  & SPM4$^{b}$ \\
\smallskip                          
B & $-3.6$  (1.2) & $+35.5$  (1.3)  & SPM4$^{b}$ \\
A & $-5.5$  (0.8) & $+19.1$  (0.8)  & UCAC4$^{c}$ \\  
\smallskip                          
B & $-3.4$  (4.5) & $+37.3$  (5.8)  & UCAC4$^{c}$ \\
A & $-6.1$  (2.1) & $+22.4$  (1.4)  & this work$^{d}$ \\
\smallskip                          
B & $-7.6$  (1.9) & $+19.6$  (1.5)  & this work$^{d}$ \\
\midrule                            
A & $-4.1$  (0.9) & $+19.9$  (0.9)  & Tycho-2$^{a}$ \\
A & $-4.6$  (0.6) & $+22.4$  (0.7)  & {\it Hipparcos}$^{e,f}$ \\
A & $-5.56$ (0.04)& $+22.15$ (0.05) & TGAS$^{e,g}$ \\
B & $-6.5$  (0.9) & $+21.8$  (0.9)  & UCAC5$^h$ \\
\bottomrule
\end{tabular}
\begin{minipage}{0.45\textwidth}
{\bf Notes:}\\
$^{a}$ {Based on very old photographic plates (AC2000) and Tycho data with a large epoch difference of about 93 years.}\\
$^{b}$ {Combining photographic plates from 1973 and CCD observations from 2007 with the Yale Southern Observatory's double-astrograph.}\\
$^{c}$ {Combining photographic plates from 1973 of Yale Southern Observatory's double-astrograph and modern UCAC CCD observations.}\\
$^{d}$ {Linear solution using the position measurements at three recent epochs only (without photographic plate measurements): Tycho \citep{esa97} from 1991, 2MASS from 1999, and {\it Gaia}  DR1 \citep{gaia16a,gaia16b} from 2015.}\\
$^{e}$ {Determined in combined proper motion and parallax solution.}\\
$^{f}$ {Revised {\it Hipparcos} solution by \citet{vanleeuwen07}.}\\
$^{g}$ {Based on Tycho-2 position from 1991 and the first 14 months of {\it Gaia} observations \citep{gaia16a,gaia16b,lindegren16}.}\\
$^{h}$ {Based on Gaia DR1 and re-reduced UCAC CCD observations using TGAS reference stars.}
\end{minipage}
\end{table}

The question whether A and B share a common proper motion had until recently no clear answer. While both stars can be found in the Tycho catalogue \citep{esa97}, only A has a proper motion in the Tycho-2 catalogue \citep{2000A&A...355L..27H}, which we list in the lower part of Table~\ref{table_pm}. But both A and B were measured on the old photographic plates used for the Astrographic Catalogue (AC2000; \citealt{urban98a}). Consequently, A and B have long-term proper motions in the ACT \citep[Astrographic Catalog/Tycho][]{urban98b} and Tycho Reference Catalogue (TRC; \citealt{hog98}). The ACT and TRC proper motions given in Table~\ref{table_pm} represent two independent reductions of the same AC2000 and Tycho data. Tycho-2 involved not only these data but also intermediate epochs from other ground-based astrometric catalogues. Probably, some of those were problematic for the proper motion solution of B. As one can see, the ACT and TRC proper motion components of A and B agree within the given errors, with B having smaller values and larger errors in declination direction.

On the other hand, the Southern Proper Motion Program (SPM4; \citealt{girard11}) and the Fourth US Naval Observatory CCD Astrograph Catalogue (UCAC4; \citealt{zacharias13}) showed a disagreement in the proper motion components of A and B in declination (see Table~\ref{table_pm}). In these two catalogues, which are only partly independent as both are based on the same first-epoch data, B has a much larger proper motion than A in declination direction. This difference is three times larger than the already relatively large errors given for B. We assume that A and B were well separated and measured with the different modern CCD observations used as second epochs for SPM4 and UCAC4, respectively. The very large and similar proper motion of B in declination may be caused by a centroiding problem on the same photographic plates of the Yale Southern Observatory double-astrograph that served as first epoch data for both catalogues. We also mention that there is a smaller discrepancy seen for the SPM4 proper motion components of A and B in right ascension, which do also not agree within their errors. 

To investigate further the questioned common proper motion of A and B, we used the presumably most accurate positional data, excluding any photographic plates, for our own proper motion measurements. The first data release of {\it Gaia}  ({\it Gaia}  DR1; \citealt{gaia16a,gaia16b,lindegren16}) provided the most accurate positions for both A and B. Due to the fact that B was not included in Tycho-2, it has unfortunately no proper motion in the Tycho-{\it Gaia} Astrometric Solution (TGAS) of {\it Gaia} DR1. However, we can combine the {\it Gaia} DR1 positions with those in the Tycho catalogue \citep{esa97} and use in addition the intermediate epoch positions given in the Two Micron All Sky Survey (2MASS; \citealt{skrutskie06}). For object A, we obtained a very similar proper motion as compared to its TGAS measurement (see Table~\ref{table_pm}). Our proper motion of B agrees within the errors with our proper motion of A. Compared to the internal agreement between A and B proper motion components in the ACT and TRC solutions, we achieved an even better agreement. Our A--B difference of the larger component in declination direction is smaller than in all proper motion solutions involving photographic plate measurements. When we included photographic positions from the AC2000 and the Southern Hemisphere Catalogue of Bordeaux \citep{rousseau96} in our proper motion determination, we arrived at values similar to those from the ACT and TRC. Our preferred results for the pair A$+$B, without using any photographic data (Table~\ref{table_pm}), are consistent with a common proper motion of A and B. Interestingly, the new UCAC5 catalogue \citep{2017arXiv170205621Z} provides an accurate proper motion of B that agrees very well with the TGAS proper motion of A (Table\,\ref{table_pm}), whereas the UCAC5 proper motion of A is unreliable (zero number of observations used for mean UCAC5 position).

\subsection{No signs of orbital motion?}
\label{SubSect_orb}

We assume that the common proper motion of A and B indicates a wide binary system. Then we can use the {\it Hipparcos} and {\it Gaia} measurements of the parallax of A to derive a distance and the projected physical separation of the system components. The TGAS parallax of A ($6.07\pm0.29$\,mas) is significantly larger than the almost identical values in the original ($4.16\pm0.63$\,mas; \citealt{esa97}) and revised ($4.17\pm0.60$\,mas; \citealt{vanleeuwen07}) {\it Hipparcos} catalogues, respectively. The formal distance is therefore reduced from about $240$\,pc according to {\it Hipparcos} to about $165$\,pc according to the TGAS. The angular separation of A and B in the TGAS is $19.388$\arcsec{}, corresponding to a projected physical separation of about $3200$~AU.

The apparent magnitudes of A and B ($V=7.604$ and $V=10.815$, respectively; \citealt{kharchenko01}) and the assumed common distance from TGAS lead to absolute magnitudes of about $M_V=1.5\pm0.1$ and $M_V=4.7\pm0.1$, respectively. These values are comparable with those of an A1 main sequence star (Sirius) and the Sun, respectively. With a spectral type of F2\,III SrEuSi: \citep{2001MNRAS.326..387K}, the $\delta$~Scuti star A is expected to be about $0.8$~mag brighter and about $1.3$ times more massive than a normal main sequence star of the same spectral type \citep{starikova79}. Therefore, it should also have a mass similar to that of Sirius. For estimating the orbital period of the system A$+$B, we assume two and one solar masses for the two components, respectively.


\begin{figure*}
\centering
\includegraphics[width=5in]{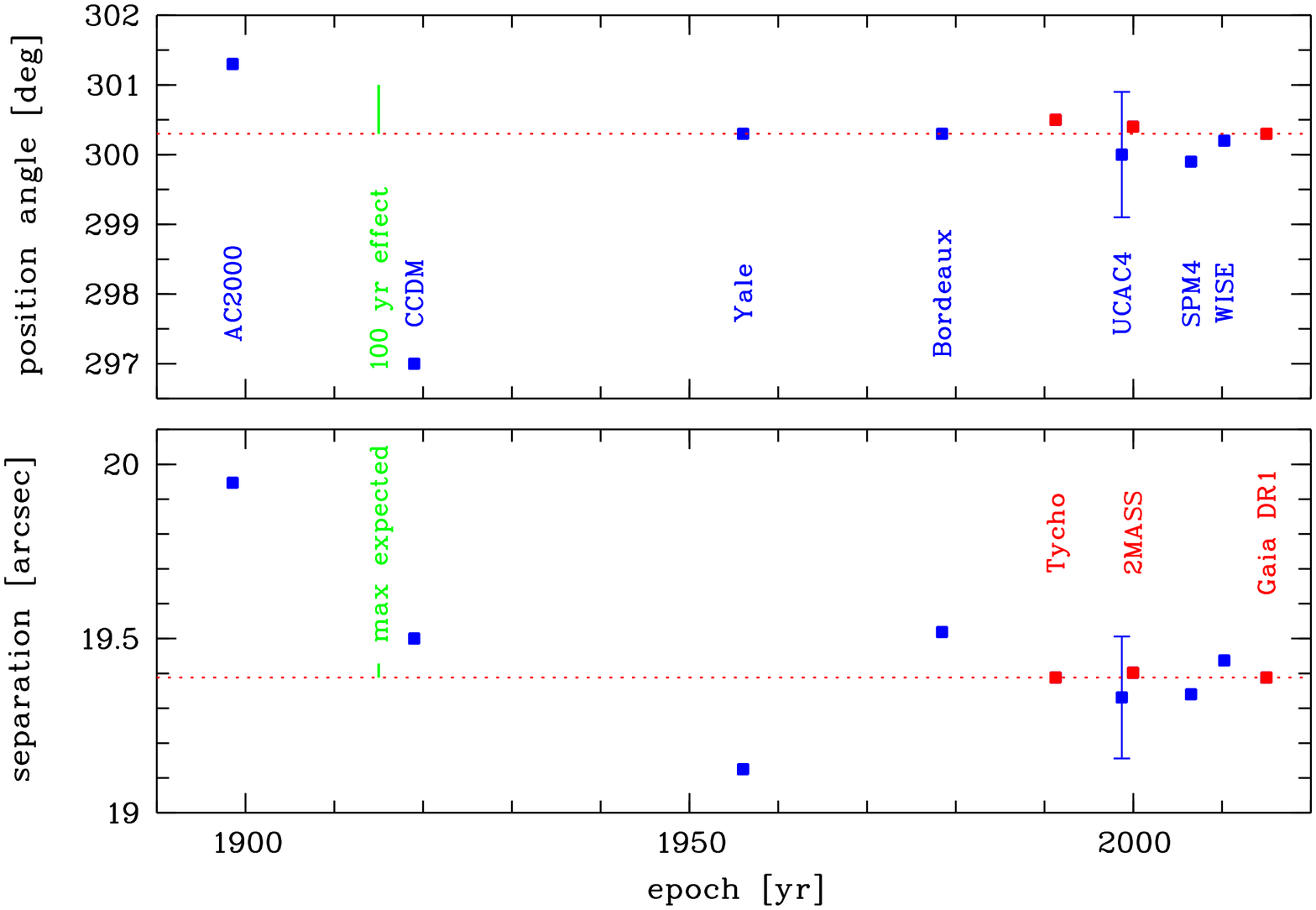}
\caption{Measured separation (bottom) and position angle (top) between A and B over time. Blue symbols and labels mark (with increasing time) data from AC2000, the Catalogue of Components of Double \& Multiple stars (CCDM; \citealt{dommanget00}), Yale observatory (\citealt{lu71} listed in \citealt{wycoff06}), Bordeaux \citep{rousseau96}, UCAC4 double stars \citep{hartkopf13}, SPM4, and the Wide-field Infrared Survey Explorer (WISE; \citealt{wright10}) all-sky catalogue. Red symbols and labels represent the most accurate data from Tycho, 2MASS, and {\it Gaia} (also indicated by the red dotted line). The green lines show the maximum expected effects over a $100$ year baseline (see text). }
\label{fig:sep_pa}
G\end{figure*}


In the simple case of a circular orbit in the plane of the sky, we estimate an orbital period of about 100\,000\,years and a differential proper motion due to orbital motion of B with respect to A of only about $2$~mas/year. This is comparable to our measured difference in the proper motion of A and B in declination but also to our proper motion measurement errors (Table~\ref{table_pm}). Over a time baseline of $100$ years, we would not yet expect to see changes in the separation and/or position angle between A and B, as these are estimated to be smaller than $40$~mas (for an edge-on circular orbit) and $0.7$\,degrees (for a circular orbit in the plane of the sky), respectively. These values are smaller than most pre-{\it Gaia} measurement errors. In Fig.~\ref{fig:sep_pa}, we show all available measurements of the separation and position angle over time (except for the probably most unreliable WDS data from 1835, which fall well outside of the plotted range for the separations and position angles). The expected largest possible effects over $100$ years are shown in green. Error bars were only available for the UCAC4 double star data \citep{hartkopf13}. They should be of similar size (or even larger at early epochs) for all other measurements, except for the presumably most accurate red data points, where they could be comparable to the symbol size. As expected, there is no clear trend but large scatter, especially at early epochs. In particular, the {\it Gaia}, 2MASS, and Tycho data, plotted in red, confirm a constant separation, whereas a constant position angle is in addition supported by the Yale and Bordeaux data.

\subsection{Space velocity}
\label{SubSect_uvw}

The tangential velocity of A is according to the TGAS $17.9\pm0.9$\,km\,s$^{-1}$, which is smaller than the radial velocity measured for B in the 5th data release of the Radial Velocity Experiment (RAVE; \citealt{2017AJ....153...75K}), $+34.0\pm1.3$\,km\,s$^{-1}$. Our own radial velocity measurements (Sect.~\ref{sect:mod}) of $+25\pm3$\,km\,s$^{-1}$ and $+19.0\pm0.5$\,km\,s$^{-1}$, respectively for A and B, are significantly smaller than the RAVE radial velocity of B. We prefer our own radial velocity measurements. The HARPS spectra with the resolution of $115,000$ have much better quality than RAVE spectra that have resolution about $7,500$. The radial velocity of the pulsating primary is not easy to determine accurately using low-resolution RAVE spectra due to the large $\vsini$ value and spectral variability caused by pulsations. These pulsations have impact on the shape and position of spectral lines. The radial velocity for the primary was calculated using the LSD (Least-Squares Deconvolution) profile in the HARPS Stokes I spectrum. The radial velocity of the secondary was estimated in a similar manner. Using the TGAS parallax measured for A for both components, our own radial velocity, and proper motion measurements for A and B, respectively, we calculated the heliocentric Galactic space velocities $UVW$ according to \citet{johnson87}. These velocity components are defined in a right-handed coordinate system so that they are positive in the direction of the Galactic centre ($U$), of Galactic rotation ($V$), and to the North Galactic Pole ($W$). All three space velocity components of the two stars, $(U,V,W)_A=(-1\pm2,-14\pm3,-28\pm2)$\,km\,s$^{-1}$ and $(U,V,W)_B=(-1\pm1,-9\pm1,-24\pm1)$\,km\,s$^{-1}$, agree to within a few km\,s$^{-1}$. The differences are comparable to or only slightly larger than the individual errors. The relatively large negative $W$-velocity and the Galactic coordinates of $(l,b)\approx(299,-32)$\,degrees indicate a current direction of motion of the pair mainly away from the Galactic plane. According to the TGAS parallax of A, the current distance from the plane is about $87$\,pc. Both kinematics and location fall in the range of the vertical velocity dispersion ($25\pm5$\,km\,s$^{-1}$) and scale height ($300\pm50$\,pc) of the (old) thin disk \citep{BlandHawthorn16}, respectively. To summarize, in spite of the high angular separation between HD\,21190 and CPD\,$-$83$^{\circ}$\,64B, the presented analysis of all available astrometric data indicates that both stars form a physical binary system.


\section{Stellar parameters and abundance determination}
\label{sect:mod}

HARPS observations were carried out on two consecutive nights: the primary HD\,21190 was observed on $2016$ June $15$, while the secondary was observed on $2016$ June $16$. The obtained observations have a signal-to-noise ratio (S/N) of about $100$ for the primary and of about $70$ for the secondary in the Stokes~$I$ spectra and a resolving power of about $R = 115\,000$. They cover the spectral range  $3780-6910$\,\AA{}, with a small gap between $5259$ and $5337$\,\AA{}. The reduction was performed using the HARPS data reduction software available at the $3.6$\,m telescope in Chile.


\begin{figure*}
\centering
\includegraphics[width=7in]{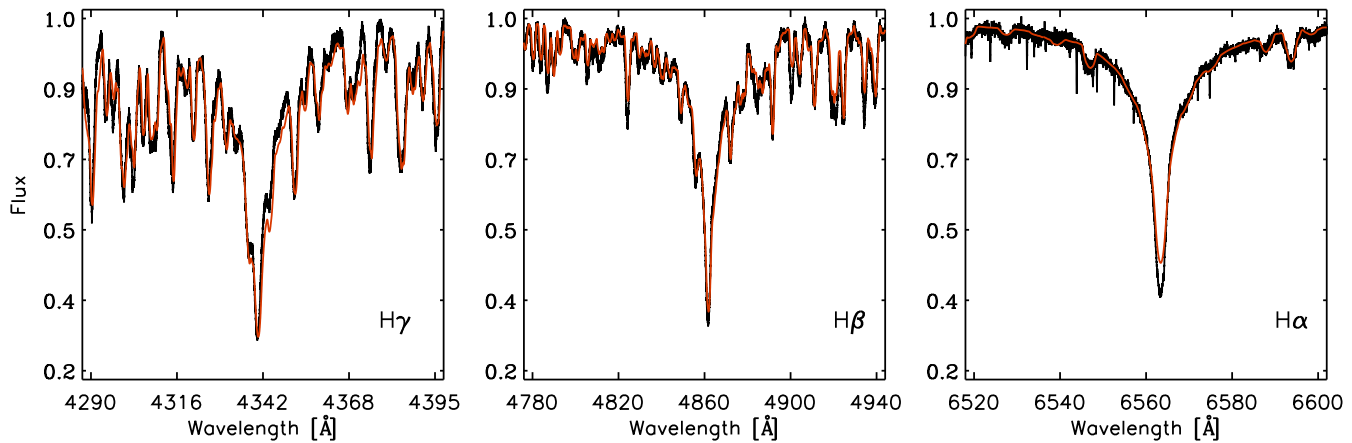}
\caption[]{Observed Balmer lines H$\gamma$, H$\beta$, and H$\alpha$ (black) and best fitted synthetic profiles (red) for HD\,21190. The sensitivity of Balmer lines to effective temperature was use to determine $\teff = 6900 \pm 100$\,K for this star. }
\label{fig:balmer_f}
\end{figure*}


\begin{figure*}
\centering
\includegraphics[width=7in]{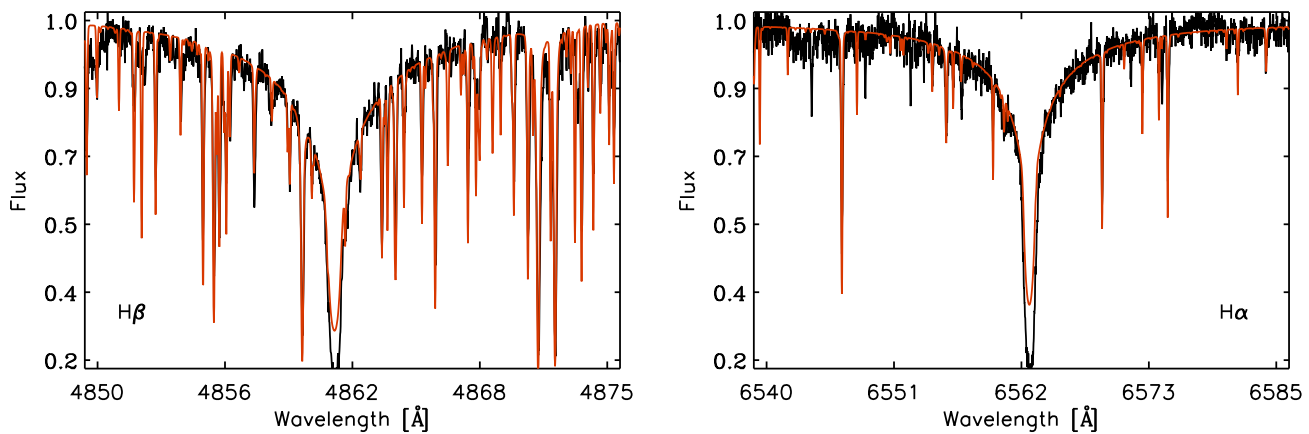}
\caption[]{Observed Balmer lines H$\beta$ and H$\alpha$  (black) and best fitted synthetic profiles (red) for CPD\,$-$83$^{\circ}$\,64B. The sensitivity of Balmer lines to effective temperature was use to determine $\teff = 5900 \pm 100$\,K for this star.}
\label{balmer_g}
\end{figure*}


\begin{figure}
\centering
\includegraphics[width=3.4in]{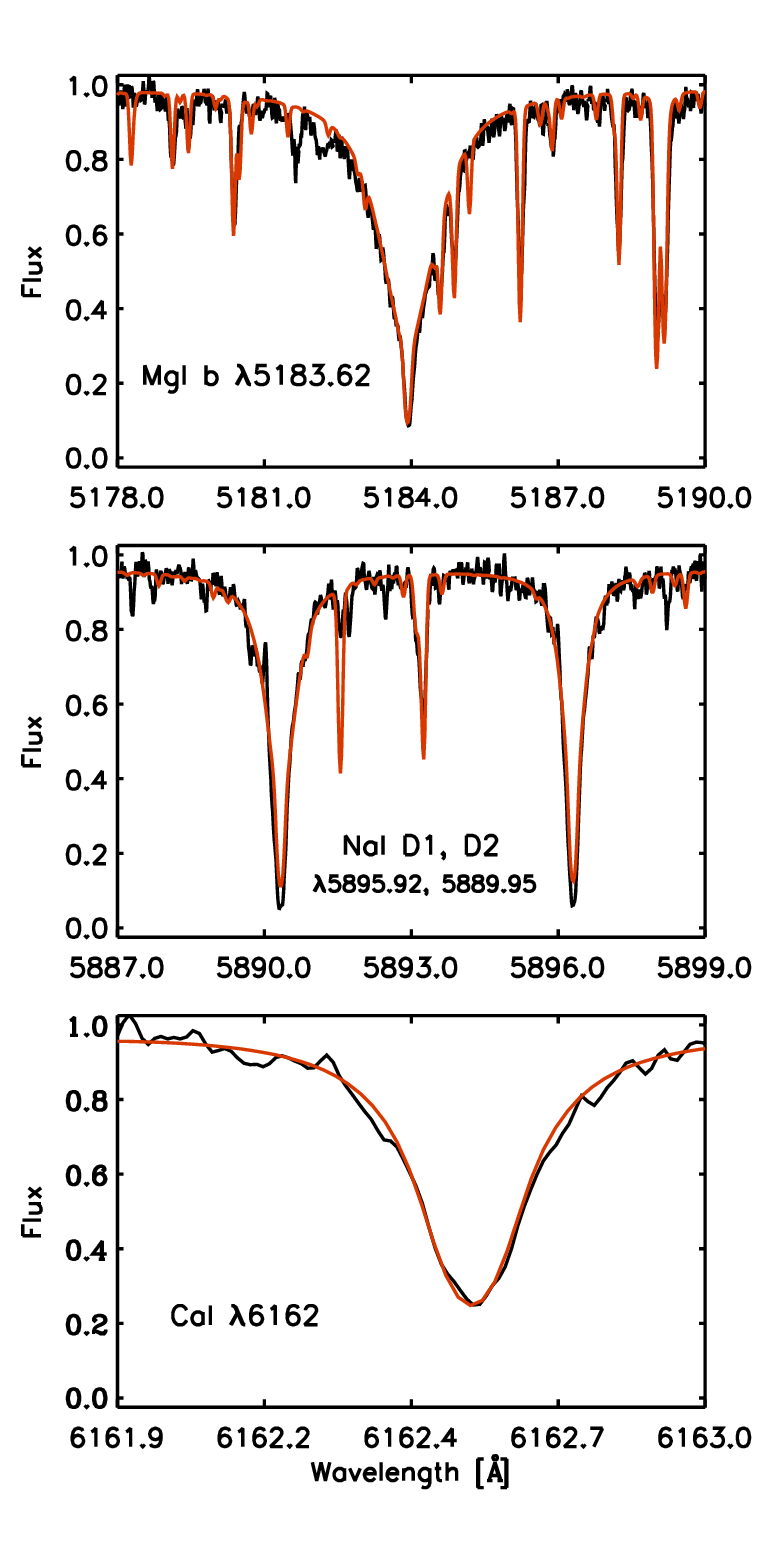}
\caption[]{Observed strong lines (black) and best fitted synthetic profiles (red) for CPD\,$-$83$^{\circ}$\,64B. Lines of \ion{Mg}{i}\,b ($5183.62$\,\AA{}), \ion{Na}{i}\,D1 ($5889.95$\,\AA{}) and D2 ($5895.92$\,\AA{}), and \ion{Ca}{i} ($6162.18$\,\AA{}) used to measure the surface gravity $4.3 \pm 0.1$ of CPD\,$-$83$^{\circ}$\,64B. }
\label{strong_g}
\end{figure}


%
Different independent sources were used to obtain stellar parameters of both components. Effective temperatures ($\teff$) and surface gravities ($\logg$) calculated from calibrations of photometric indices and from the spectral energy distribution (SED) were treated as initial parameters for the spectroscopic analysis. Further, a synthesis of hydrogen line profiles as well as the requirement of ionization and excitation equilibria in the Fe lines were used to get $\teff$, $\logg$, and microturbulence ($\micro$) for the F-type primary whereas for the G-type secondary CPD\,$-$83$^{\circ}$\,64B the atmospheric parameters were obtained from Balmer lines, strong \ion{Na}{i}\,D1 ($5889.95$\,\AA{}) and D2 ($5895.92$\,\AA{}), \ion{Ca}{i} ($6162.18$\,\AA{}), \ion{Mg}{i}\,b ($5183.62$\,\AA{}), and \ion{Fe}{i} lines.

To calculate the necessary synthetic spectra, we used atmospheric models (plane-parallel, hydrostatic and radiative equilibrium, 1-dimensional) and synthetic spectra computed with the line-blanketed, local thermodynamical equilibrium (LTE) {\small\sc ATLAS\,9} code \citep{1993KurCD..13.....K}. The grid of atmospheric models was calculated for effective temperatures from $5000$ to $6200$\,K with a step of $50$\,K, and from $6200$ to $12000$\, with a step of 100\,K. Surface gravities from $2.0$ to $4.6$\,dex with a step of $0.1$\,dex, microturbulence velocities between $0.0$ and $6.0$\,km\,s$^{-1}$ with a step of $0.1$\,km\,s$^{-1}$, and metallicities [M/H] equal to $0.0$, $+0.5$ and $+1.0$ were considered for each temperature. Appropriate synthetic spectra were computed with the {\small\sc SYNTHE} code \citep{1993KurCD..18.....K}. Both codes, {\small\sc ATLAS9} and {\small\sc SYNTHE}, were ported to GNU/Linux by \citet{sbordone}. The line list used was that of Robert Kurucz\footnote{http://kurucz.harvard.edu} \citep{2014dapb.book...63K}.

\subsection{Atmospheric parameters of the primary component HD\,21190}

For HD\,21190, Str\"{o}mgren {\it uvby$\beta$} \citep{1998A&AS..129..431H}, Johnson $BV$ \citep{2000A&A...355L..27H}, and 2MASS $JHK$ \citep{2003yCat.2246....0C} photometry is available and was used to determine the initial parameters.

The $uvby\beta$ indices were taken from the General Catalogue of Photometric Data (GCPD) \citep{1998A&AS..129..431H}\footnote{http://obswww.unige.ch/gcpd/gcpd.html}. Atmospheric parameters were determined using the {\small\sc UVBYBETA} code written by \citet{1985MNRAS.217..305M} and corrected by \citet{1993A&A...268..653N}. We obtained the effective temperature $6900 \pm 100$\,K and surface gravity $3.63 \pm 0.17$\,dex. The errors of $\teff$ and $\logg$ were calculated taking into account the uncertainties of the photometric indices and the internal error of the method.

To calculate atmospheric parameters from the $(V-K)$ index, knowledge of the interstellar reddening parameter is necessary. The colour excess $E(B-V) = 0.005 \pm 0.003$\,mag was derived using the relation between the equivalent width of the Na\,D$_{2}$ line ($5889.95$\,\AA{}) and the $E(B-V)$\,mag \citep{1997A&A...318..269M}. The $(V-K)$ index was determined on the basis of the $K$ magnitudes from the 2MASS catalogue \citep{2003yCat.2246....0C} and the $V$ magnitudes from the GCPD catalogue. It was used to determine $\teff$ from the \citet{2006A&A...450..735M} relation. We used the relation $E(V-K) = 2.72 E(B-V)$ from \citet{1979ARA&A..17...73S}. In the calculations we assumed $\rm{[Fe/H]} = 0.0$ and $\log g = 4.0$. This calibration has only a weak dependence on $\logg$ and metallicity. The obtained effective temperature is equal to $6830 \pm 70$\,K. We determined the uncertainty of $\teff$ by changing $\logg$ by $0.2$ and [Fe/H] by $0.2$\,dex.

The stellar effective temperature was also determined from the spectral energy distribution (SED), constructed from the photometry used before ($V$, $uvby$, $2MASS$) supplemented by $B$, $U$, $R$ and $I$ magnitudes adopted from GCPD database. Effective temperatures were determined by fitting \citet{1993KurCD..13.....K} model fluxes to the de-reddened SED. The metallicity [M/H] is poorly constrained by our SED, therefore we fixed [M/H]\,$=0.0$\,dex. The model fluxes were convolved with photometric filter response functions. From this method, we obtained an effective temperature $7100 \pm 200$\,K and $\logg = 3.70 \pm 0.20$. The uncertainties of the derived parameters were estimated taking into account the errors of the photometric colours, $E(B-V)$, and the assumed error $0.1$ for [M/H].


The value of $\teff = 6900 \pm 100$\,K was obtained using the sensitivity of the Balmer H$\alpha$, H$\beta$ and H$\gamma$ lines to effective temperature. Hydrogen lines are not good indicators of $\logg$ for effective temperatures lower than about $8000$\,K \citep{2014dapb.book...85S}. Therefore, the surface gravity was assumed to be equal to $4.0$\,dex. We calculated synthetic Balmer lines assuming solar metallicity. We used an iterative approach to minimize the differences between the observed and the synthetic Balmer profiles \citep[see][]{2004A&A...425..641C}. The uncertainties of the derived values were estimated taking into account uncertainties resulting from the validity of the normalisation. Since hydrogen lines in high-resolution HARPS spectra extend over adjacent orders, to improve the normalization, we also used recent FORS\,2 low-resolution observations of HD\,21190. To estimate the error of $\teff$, we took into account the differences in the determined $\teff$ values from the separate Balmer lines. In Fig.~\ref{fig:balmer_f} we present the observed Balmer lines together with the synthetic profiles.

The photometric methods, SED, and Balmer line analysis provided input atmospheric parameters $\teff$ and $\logg$ for the iron line investigation. In the case of HD\,21190, the analysis of individual lines was impossible because of relatively high rotation velocity, hence we assumed that the iron abundances obtained from different spectral regions are the same for the correct $\teff$, $\logg$ and $\micro$ values. We were looking for $\teff$ in the range $6700-7200$\,K, $\logg$ from $3.0$ to $3.8$, and $\micro$ from $1.5$ to $3.5$\,\kms{}. The final effective temperature $6900 \pm 100$\,K, surface gravity $3.6 \pm 0.2$\,dex and microturbulence $2.4 \pm 0.2$\,\kms{} were used for the elemental composition determination of HD\,21190. A moderate $\vsini = 74 \pm 3$\,\kms{} was measured in the HARPS spectrum.

\begin{table*}
\caption{Atmospheric parameters of HD\,21190 (A) and CPD\,$-$83$^{\circ}$\,64B (B) determined from photometric and spectroscopic methods (see the first column).} 
\label{table_atm}             
\centering                   
\begin{tabular}{ll}       
\toprule
\multicolumn{2}{c}{HD\,21190} \\
\midrule
Str\"{o}mgren {\it uvby$\beta$}          &$\teff = 6900 \pm 100$\,K, $\logg = 3.63 \pm 0.17$                \\
2MASS $(V-K)$ index                      &$\teff = 6830 \pm 70$\,K; $\logg$ and [Fe/H] assumed              \\
SED                                      &$\teff = 7100 \pm 200$\,K, $\logg = 3.70 \pm 0.20$, [M/H] assumed \\
Balmer H$\alpha$, H$\beta$ and H$\gamma$ &$\teff = 6900 \pm 100$\,K; $\logg$ and [M/H] assumed              \\
Fe lines                                 &$\teff = 6900 \pm 100$\,K, $\logg = 3.60 \pm 0.20$,               \\
                                         &$\xi_{\rm t} = 2.4 \pm 0.2$\,\kms, $\vsini = 74 \pm 3$\,\kms\     \\
\midrule
\multicolumn{2}{c}{CPD\,$-$83$^{\circ}$\,64B} \\
\midrule
Balmer H$\beta$ and H$\alpha$                                              & $\teff = 5900 \pm 100$\,K; $\logg$ and [M/H] assumed \\
Strong lines: \ion{Na}{i}\,D1 ($5889.95$\,\AA{}) and D2 ($5895.92$\,\AA{}),& $\logg = 4.3 \pm 0.1$, $\teff$, $\micro$, and [M/H] assumed     \\
\ion{Ca}{i} ($6162.18$\,\AA{}), and \ion{Mg}{i}\,b ($5183.62$\,\AA{})      &    \\
Fe lines                                                                   & $\teff = 5850 \pm 50$\,K, $\micro = 1.0\pm0.2$\,\kms, \\
                                                                           & $\vsini{} = 3.1 \pm 0.7$\,\kms; $\logg$ assumed   \\
\bottomrule
\end{tabular}
\end{table*}


\begin{table*}
\centering
\caption{Abundances of chemical elements for both components. An average value is given; the standard deviations were calculated only if the number of analysed lines is greater than two. The number of the analysed lines is also indicated.}
\label{tab:abundances}
\begin{tabular}{lrcrcccrcccc}
\toprule
El. &Atomic & \vline & No. of & Average   & Standard  & \vline & No. of & Average   & Standard  & \vline & Solar  \\
    &number & \vline & lines  & abundance & deviation & \vline & lines  & abundance & deviation & \vline & abundances$^*$ \\
\midrule
 & & \vline & \multicolumn{3}{c}{HD\,21190}&  \vline & \multicolumn{3}{c}{CPD\,$-$83$^{\circ}$\,64B} & \vline \\
\midrule
 C  & $ 6$  & \vline & $ 7 $& $ 8.67 $& $ 0.12 $ & \vline &$   1  $&$  8.71 $&$  -   $ & \vline & $8.43\pm0.05$  \\
 Na & $11$  & \vline & $ 3 $& $ 6.60 $& $ 0.28 $ & \vline &$   9  $&$  6.42 $&$  0.22$ & \vline & $6.21\pm0.04$  \\
 Mg & $12$  & \vline & $ 7 $& $ 7.81 $& $ 0.37 $ & \vline &$   6  $&$  7.62 $&$  0.18$ & \vline & $7.59\pm0.04$  \\
 Si & $14$  & \vline & $14 $& $ 7.68 $& $ 0.21 $ & \vline &$  24  $&$  7.58 $&$  0.25$ & \vline & $7.51\pm0.03$  \\
 S  & $16$  & \vline & $ 4 $& $ 7.61 $& $ 0.01 $ & \vline &$   1  $&$  7.28 $&$  -   $ & \vline & $7.12\pm0.03$  \\
 Ca & $20$  & \vline & $13 $& $ 6.94 $& $ 0.22 $ & \vline &$  36  $&$  6.51 $&$  0.21$ & \vline & $6.32\pm0.03$  \\
 Sc & $21$  & \vline & $ 7 $& $ 3.69 $& $ 0.19 $ & \vline &$   8  $&$  3.10 $&$  0.05$ & \vline & $3.16\pm0.04$  \\
 Ti & $22$  & \vline & $29 $& $ 5.21 $& $ 0.25 $ & \vline &$ 124  $&$  4.90 $&$  0.16$ & \vline & $4.93\pm0.04$  \\
 V  & $23$  & \vline & $ 4 $& $ 4.39 $& $ 0.24 $ & \vline &$  27  $&$  3.92 $&$  0.16$ & \vline & $3.89\pm0.08$  \\
 Cr & $24$  & \vline & $32 $& $ 5.98 $& $ 0.18 $ & \vline &$ 134  $&$  5.68 $&$  0.15$ & \vline & $5.62\pm0.04$  \\
 Mn & $25$  & \vline & $ 8 $& $ 5.09 $& $ 0.21 $ & \vline &$  39  $&$  5.36 $&$  0.15$ & \vline & $5.42\pm0.04$  \\
 Fe & $26$  & \vline & $ 80$& $ 7.71 $& $ 0.13 $ & \vline &$ 599  $&$  7.56 $&$  0.16$ & \vline & $7.47\pm0.04$  \\
 Co & $27$  & \vline & $ 3 $& $ 5.73 $& $ 0.15 $ & \vline &$  27  $&$  4.94 $&$  0.18$ & \vline & $4.93\pm0.05$  \\
 Ni & $28$  & \vline & $44 $& $ 6.82 $& $ 0.16 $ & \vline &$ 134  $&$  6.30 $&$  0.17$ & \vline & $6.20\pm0.04$  \\
 Cu & $29$  & \vline & $ 2 $& $ 4.14 $& $ -    $ & \vline &$   3  $&$  4.13 $&$  0.12$ & \vline & $4.18\pm0.05$  \\
 Zn & $30$  & \vline & $ 2 $& $ 4.46 $& $ -    $ & \vline &$   3  $&$  4.76 $&$  0.08$ & \vline & $4.56\pm0.05$  \\
 Ge & $32$  & \vline & $   $& $      $& $      $ & \vline &$   1  $&$  3.84 $&$   -  $ & \vline & $3.63\pm0.07$  \\
 Sr & $38$  & \vline & $ 1 $& $ 3.97 $& $ -    $ & \vline &$   2  $&$  3.54 $&$   -  $ & \vline & $2.83\pm0.06$  \\
 Y  & $39$  & \vline & $ 7 $& $ 3.08 $& $ 0.42 $ & \vline &$  12  $&$  2.58 $&$  0.22$ & \vline & $2.21\pm0.05$  \\
 Zr & $40$  & \vline & $ 3 $& $ 3.16 $& $ 0.27 $ & \vline &$   4  $&$  2.69 $&$  0.25$ & \vline & $2.59\pm0.04$  \\
 Ru & $44$  & \vline & $   $& $      $& $      $ & \vline &$   1  $&$  1.91 $&$  -   $ & \vline & $1.75\pm0.08$  \\
 Ba & $56$  & \vline & $ 2 $& $ 3.74 $& $ -    $ & \vline &$   5  $&$  2.53 $&$  0.12$ & \vline & $2.25\pm0.07$  \\
 La & $57$  & \vline & $ 3 $& $ 2.09 $& $ 0.17 $ & \vline &$   4  $&$  1.44 $&$  0.11$ & \vline & $1.11\pm0.04$  \\
 Ce & $58$  & \vline & $ 1 $& $ 3.30 $& $ -    $ & \vline &$   7  $&$  1.71 $&$  0.25$ & \vline & $1.58\pm0.04$  \\
 Nd & $60$  & \vline & $13 $& $ 2.29 $& $ 0.26 $ & \vline &$  24  $&$  1.70 $&$  0.27$ & \vline & $1.42\pm0.04$  \\
 Sm & $62$  & \vline & $   $& $      $& $      $ & \vline &$  10  $&$  1.69 $&$  0.28$ & \vline & $0.95\pm0.04$  \\
 Eu & $63$  & \vline & $ 1 $& $ 1.69 $& $ -    $ & \vline &$   1  $&$  0.60 $&$  -   $ & \vline & $0.52\pm0.04$  \\
 Dy & $66$  & \vline & $   $& $      $& $      $ & \vline &$   1  $&$  1.87 $&$  -   $ & \vline & $1.10\pm0.04$  \\
\bottomrule
\end{tabular}
\begin{minipage}{\textwidth}
{\bf Notes:}\\
$^*$ Solar abundances:
C -- \citet{2009ARA&A..47..481A};
Na--Ca -- \citet{2015A&A...573A..25S};
Sc--Ni -- \citet{2015A&A...573A..26S};
Cu--Dy -- \citet{2015A&A...573A..27G}.
\end{minipage}
\end{table*}


\begin{figure*}
\includegraphics[width=7in]{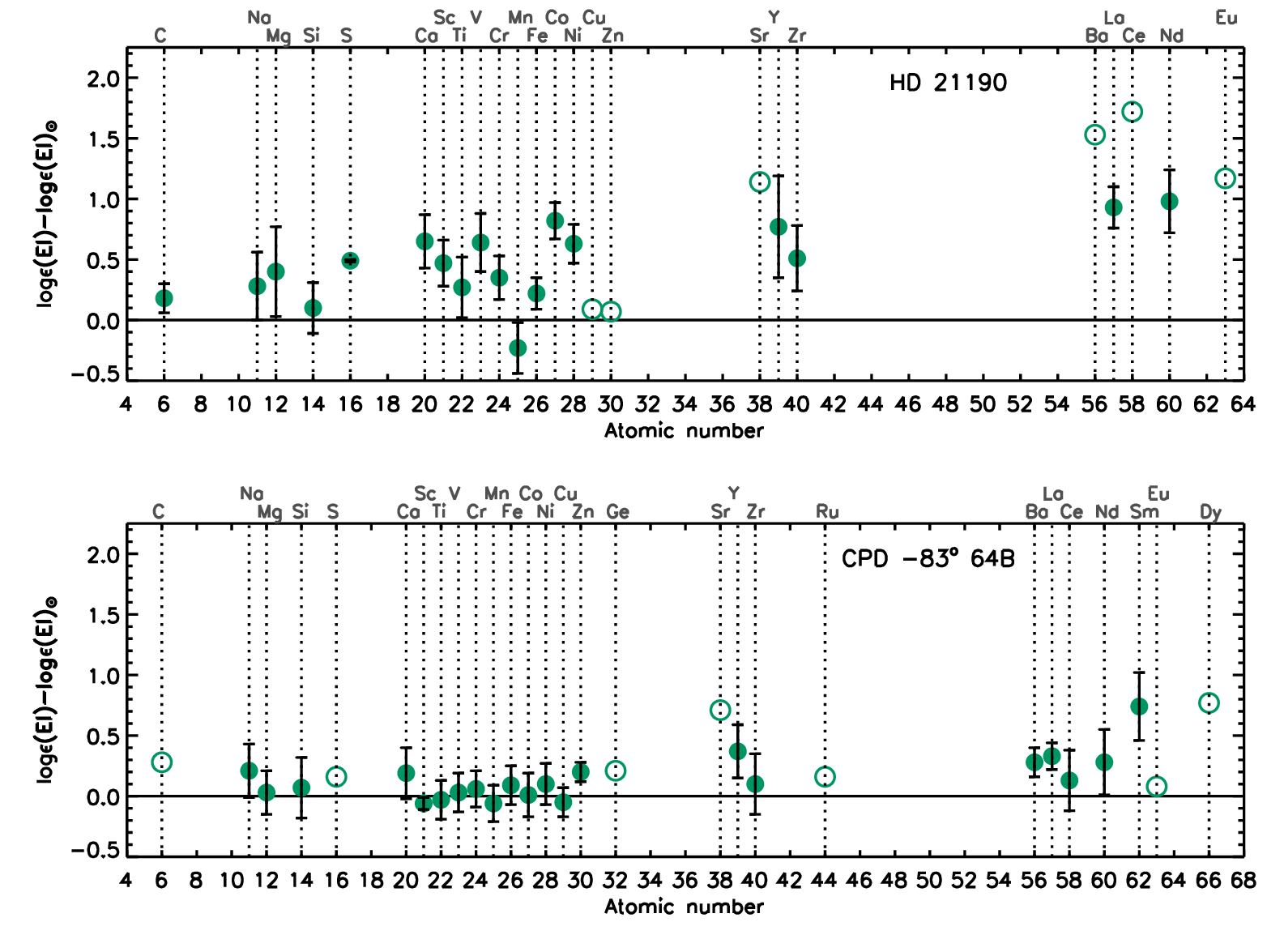}
\caption[]{Chemical abundances of HD\,21190 (upper panel) and CPD\,$-$83$^{\circ}$\,64B (lower panel) compared to solar values, $\log\epsilon({\rm El})-\log\epsilon({\rm El})_\odot$ (see Table~\ref{tab:abundances}). The values determined from more than two lines are shown as filled circles, whereas those with less lines are denoted as open circles. The error bars are standard deviations, calculated in cases where at least three lines or blends were used to determine the abundance of the given element.}
\label{abundances-f}
\end{figure*}


The summary of obtained atmospheric parameters is shown in Table\,\ref{table_atm}. We determined consistent results from every method we use for the atmospheric parameter determination. The $\logg$ parameter is more problematic, as only the {\small\sc UVBYBETA} code and SED analysis give values for the surface gravity. In comparison with the $uvby\beta$ photometry, the SED analysis gives similar value of $\logg = 3.70$. To determine the proper value of $\logg$ from the last method, a well constructed SED is necessary, especially in the region of the Balmer jump. The value $3.60$\,dex obtained from the iron line analysis is consistent with the $\logg$ from the {\small\sc UVBYBETA} code and with the SED investigation.

The obtained atmospheric parameters of HD\,21190 are also consistent with the values obtained in a few previous studies. \citet{2012MNRAS.427..343M} used various surveys to fit the SED and obtained an effective temperature of $6817$\,K. \citet{2011A&A...530A.138C} give $\teff = 7030 \pm 120$\,K derived via the InfraRed Flux Method, and $\logg = 3.49$, computed from the fundamental definition involving $\teff$, the bolometric flux, and mass. \citet{2001MNRAS.326..387K} derived an effective temperature $6780 \pm 100$\,K and a surface gravity $3.34 \pm 0.15$\,dex using a multidimensional downhill simplex technique developed by \citet{2001AJ....121.2159G}, that simultaneously finds the best fit to the observed spectrum and fluxes determined from photometry from a grid of synthetic spectra and fluxes computed using model atmospheres from the stellar atmosphere program {\small\sc ATLAS\,9} \citep{1993KurCD..13.....K}.

\subsection{Atmospheric parameters of the secondary component CPD\,$-$83$^{\circ}$\,64B}

Only Johnson $BV$ and 2MASS $JHK$ colours are available for the secondary component CPD\,$-$83$^{\circ}$\,64B. These photometries can be used only for effective temperature determination, and require the knowledge of colour excess, surface gravity, and metallicity. Therefore we decided to use various spectral lines to determine atmospheric parameters of secondary component.

A value of $\teff = 5900 \pm 100$\,K was obtained using the sensitivity of the Balmer H$\beta$ and H$\alpha$ lines to the effective temperature (see Fig.\,\ref{balmer_g}). The surface gravity was assumed to be equal to $4.0$\,dex, as this parameter has no influence on this analysis. The uncertainties of the derived values were estimated taking into account the uncertainties resulting from the validity of the normalisation. We calculated them taking into account the differences in the determined $\teff$ values from the separate Balmer lines. 

Strong lines like \ion{Na}{i}\,D1 ($5889.95$\,\AA{}) and D2 ($5895.92$\,\AA{}), \ion{Ca}{i} ($6162.18$\,\AA{}), and \ion{Mg}{i}\,b ($5183.62$\,\AA{}) show strong pressure-broadened wings in the spectra of cool stars, and can be used for $\logg$ determination \citep{gray}. However, the influence of the effective temperature and the abundances have to be taken into account. We assumed $\teff = 5900\pm100$\,K as obtained from the Balmer lines, solar metallicity [M/H], and a microturbulence $\micro = 1.0\pm0.2$\,\kms{}. A surface gravity $4.3 \pm 0.1$ was determined from the strong lines mentioned before. The quality of the fitting can be seen in Figure\,\ref{strong_g}. The uncertainty of $\logg$ was determined by changing the assumed $\teff$, [M/H], and $\micro$ in their error bars.

The final $\logg$ was adopted from the strong line analysis, but the final $\teff$ was determined from the dependency of the iron abundance on the excitation potential of the \ion{Fe}{i} lines. Analysing the iron lines, we proceeded according to the following scheme: first, $\micro$ was adjusted until there was no correlation between iron abundances and line depths for the \ion{Fe}{i} lines; second, $\teff$ was changed until there was no trend in the iron abundance versus excitation potential for the \ion{Fe}{i} lines. The obtained value $5850 \pm 50$\,K is consistent with those from the Balmer line investigation and $\micro = 1.0\pm0.2$\,\kms{} is typical for G type stars. Simultaneously, we obtained $\vsini{} = 3.1 \pm 0.7$\,\kms{}.


\subsection{Chemical abundances of both components}
\label{subsect:chem}

For both stars, the analysis of all available metal lines lets us derive their chemical compositions. We analysed the metal lines using the spectrum synthesis method, relying on an efficient least-squares optimisation algorithm \citep[for more information see][]{2015MNRAS.450.2764N}. The chosen method allows us a simultaneous determination of various parameters influencing the stellar spectra. The synthetic spectrum depends on $\teff$, $\logg$, $\micro$, $\vsini$, and the relative abundances of the elements. All these parameters are correlated, hence $\teff$, $\logg$, and $\micro$ were determined before the chemical abundance analysis. The $\vsini$ values were determined by comparing the shapes of the observed metal line profiles with the computed profiles, as shown by \citet{gray}. Chemical abundances and $\vsini$ values were determined from many different individual lines or line parts (in case of strong blending) of the spectrum. We took into account all elements that showed lines in the analysed spectral region. Elements that have little or no influence in a given part are assumed to have solar abundances. At the end, we derived the average values of $\vsini$ and abundances of all chemical elements considered for a given star.


\begin{figure}
\includegraphics[width=3.5in]{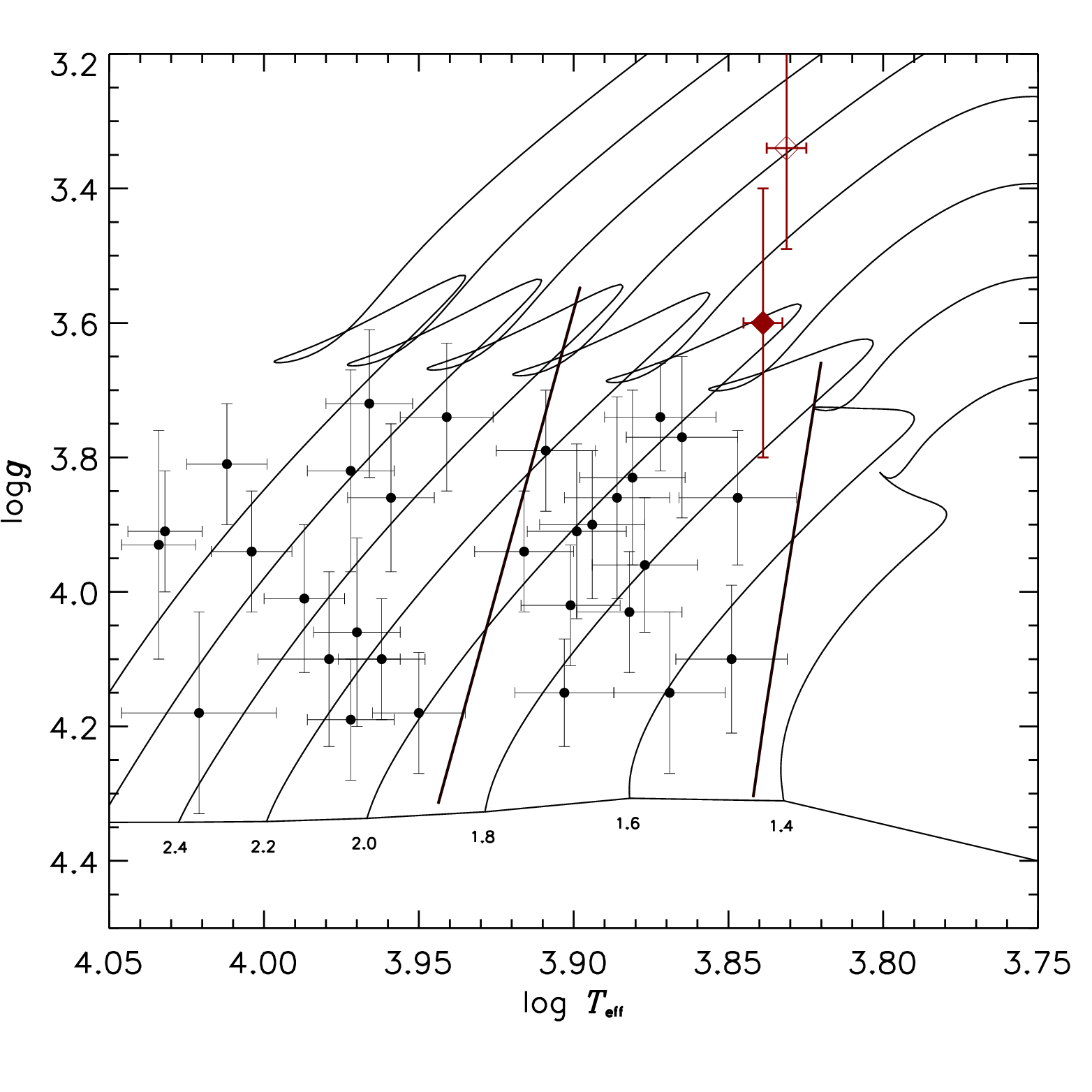}
\caption[]{Position of HD\,21190 (filled red symbol) on the $\log\teff - \logg$ diagram. Atmospheric parameters determined in this work are shown with red filled diamod, whereas those taken from \citet{2001MNRAS.326..387K} are plotted with red open diamond. Magnetic Ap stars (black filled circles) discussed in \citet{2000ApJ...539..352H} are shown for comparison. Evolutionary tracks used were taken from \citet{2015MNRAS.450.2764N}. They were calculated with Time Dependent Convection \citep[TDC,][]{2005A&A...434.1055G}, for stars with solar metallicity and $\alpha_{\rm MLT} = 1.8$. The instability strip for $\delta$\,Sct is indicated by black solid lines \citep{2000ASPC..210..215P}.}
\label{hr-diagram}
\end{figure}


In Table~\ref{tab:abundances}, we present the average abundances of both stars together with the solar values of \citet{2009ARA&A..47..481A}, \citet{2015A&A...573A..25S}, \citet{2015A&A...573A..26S}, and \citet{2015A&A...573A..27G}. In Fig.\,\ref{abundances-f} the determined average abundances are compared with the solar abundances. For the primary star, the abundance analysis was carried out for $24$ elements, while abundances for $28$ elements were studied in the spectrum of the secondary.
 
\citet{2001MNRAS.326..387K} compared a spectrum of HD\,21190 with those of well-known stars and MK standards and classified this star as a chemically peculiar F2\,III\,SrEuSi: object due to the enhancement of the spectral lines \ion{Sr}{ii}\,$4077$\,\AA{}, \ion{Eu}{ii}\,$4205$\,\AA{}, and the \ion{Si}{ii} doublet at $4128-4131$\,\AA{}. In our analysis, all studied heavy and rare-earth elements appear overabundant, confirming the peculiar nature of HD\,21190.

In contrast to the results of \citet{2001MNRAS.326..387K}, the silicon abundance obtained using fourteen lines in the HARPS spectrum is only slightly higher than the solar value. Because of strong blending, we were not able to analyse lines at wavelengths shorter than $4150$\,\AA{}. 

The most accurate abundances were derived for elements showing a rich line spectrum. In the case of HD\,21190, these are the elements Si and Ca, and the iron-peak elements Ti, Cr, Fe, and Ni. The iron-peak elements Ti, Cr, and Fe show only slightly higher abundances than solar values, while calcium and nickel are definitely overabundant. The heavy elements Y and Zr, as well as the rare-earth elements are significantly overabundant. Of all these elements, only Y and Nd show enough number of lines for an reliable abundance analysis.

The spectrum of CPD\,$-$83$^{\circ}$\,64B was not analysed before. Since the projected rotation velocity of this star is low, it was possible to carry out an abundance analysis for $28$ elements. With respect to elements with low atomic numbers, Si shows solar abundance, whereas the other considered elements are slightly overabundant. Most lines in the spectrum represent iron-peak elements, from Ca to Cu. All these elements have abundances close to the solar values. All considered heavy and rare-earth elements appear overabundant, but show larger uncertainties. Besides, only for Y, Nd, and Sm ten or more lines were analysed, for the other considered elements only few lines or blends were available. In fact Zr, Ce, and Nd can be solar in the error bars.

To check the reliability of our abundance determinations, we analysed a high S/N day sky solar spectrum taken with HARPS, with a resolving power of about $115\,000$, and covering the spectral range $3780-6860$\,\AA{}. The detailed description of this check is provided in the Appendix.


\section{Discussion}
\label{sect:disc}

Based on a comprehensive astrometric analysis, we show that the stars HD\,21190 and CPD\,$-$83$^{\circ}$\,64B most probably form a physical binary system. The proper motion of CPD\,$-$83$^{\circ}$\,64B agrees within errors with the proper motion of HD\,21190. Using radial velocity values measured in the HARPS spectra and the TGAS parallax measured for HD\,21190, we calculated for both components the heliocentric Galactic space velocities, which agree within a few km\,s$^{-1}$. With our estimated physical separation of a few thousand AU, this wide binary falls well in the range of the new TGAS wide binaries sample of \citet{2017arXiv170407829A}, with similarly small proper motions and relatively bright magnitudes. According to their discussion, our wide binary components are on the one hand not expected to interact over their lifetimes but their separation is also far below the 'tidal limit', where they could not remain gravitationally bound for long in the Galactic tidal field.

Available photometric databases and high-resolution spectroscopic data were used to derive the atmospheric parameters and chemical abundances of both components. For normal stars with similar spectral types nearly solar element abundances are expected. The detection of overabundances of several elements suggest the presence of chemical peculiarities in the atmospheres of HD\,21190. The obtained abundance pattern presented in Table~\ref{tab:abundances} is in good agreement with the previous results of \citet{2001MNRAS.326..387K}, who indicated the presence of chemical peculiarities in HD\,21190.

Two classes of chemically peculiar (CP) stars, the so-called metallic line (Am) stars and magnetic Ap stars, populate the same region in the H-R diagram. The Am stars exhibit overabundances of most iron-peak elements and some heavy elements such as Zn, Sr, Zr, and Ba, and underabundances of Ca and Sc \citep{2009ssc..book.....G}. The majority of Am stars are members of close binary systems.  Chemically peculiar magnetic Ap stars show slight overabundances of iron-peak elements, and stronger overabundances of Sr, Y, Zr, and rare-earth elements (e.g.\ \citealt{2005MNRAS.356.1256C}). The coolest members of the Ap stars, the roAp stars with pulsating periods of $6$ to $24$\,min have effective temperatures in the range of about $6600$ to $8000$\,K \citep{2006MNRAS.370.1274K}. The comparison of the results of our abundance analysis of HD\,21190 listed in Table~\ref{tab:abundances} with diverse abundance analyses presented in the literature indicates a better match with the abundances published for magnetic Ap stars. Whether HD\,21190 can be considered a roAp star is not completely clear yet: \citet{2008MNRAS.384.1140G} obtained for this star $14$ UVES spectra covering about $50$\,min, but no spectral features moving at higher frequencies were identified in the behaviour of the line profiles.

The comparison of the determined atmospheric parameters of HD\,21190, $\teff = 6900\pm100$\,K and $\logg = 3.6\pm0.2$, with these of typical magnetic Ap stars presented in Tables\,1 and 2 in the work by \citet{2000ApJ...539..352H} indicates that this star is an evolved magnetic Ap star, even with $\logg$ suggesting less evolved star than it was stated by \citet{2001MNRAS.326..387K}. The position of HD\,21190 in the $\teff - \logg$ diagram is shown on Fig.\,\ref{hr-diagram}, together with the positions of Ap stars analysed by \citet{2000ApJ...539..352H}. Evolutionary tracks used were taken from the paper of \citet{2015MNRAS.450.2764N}. They were calculated with Time Dependent Convection \citep[TDC,][]{2005A&A...434.1055G}, for stars with solar metallicity, $\alpha_{\rm MLT} = 1.8$, and cover a range of masses from $1.4$ to $2.8$\,M$_\odot$. The instability strip for $\delta$\,Sct is shown as well \citep{2000ASPC..210..215P}. HD\,21190 is located close to the Terminal Age Main-Sequence and it's position shows that it is the most evolved Ap star among the considered objects, however the error of surface gravity is relatively large. Moreover, the star is located well inside the instability strip of the $\delta$\,Scuti pulsators.


\section*{Acknowledgments}
EN acknowledges the Polish National Science Center grant no. 2014/13/B/ST9/00902. 
Calculations have been carried out at the Wroc{\l}aw Centre for Networking and Supercomputing (http://www.wcss.pl), grant No.\,214.
Based on observations made with ESO Telescopes at the La Silla Paranal Observatory under programme IDs~097.C-0277 and 097.D-0478.



\appendix
\label{appendixA}
\section{Solar abundances}

To check the reliability of our abundance determinations, we analysed a high S/N day sky solar spectrum taken with HARPS, with a resolving power of about $115\,000$, and covering the spectral range $3780-6860$\,\AA{}. The spectrum is a combination of about $30$ spectra reaching a total S/N greater than $1000$ and was added to the HARPS Solar Spectra Collection\footnote{www.eso.org/sci/facilities/lasilla/instruments/harps} by \citet{2006A&A...454..341D}.

A solar atmospheric model was calculated with Kurucz's code {\small\sc ATLAS\,9} for $T_{\rm eff} = 5777$\,K, $\log g = 4.4$, and $\xi = 1.0$\,\kms{}. The synthetic spectra were calculated using the {\small\sc SYNTHE} code, assuming macroturbulence $\zeta_t = 3.20$\,\kms{} \citep{2014MNRAS.444.3592D}. We considered all metal lines visible in the analysed solar spectrum and for which atomic data are available in Kurucz's line list. The obtained abundances are presented in Table~\ref{tab:solar}. For chemical elements listed in the first two columns, we determined abundances from all lines found in the analysed spectrum (Columns 3, 4, and 5 in Table~\ref{tab:solar}). Only some of these lines were used in \citet{2015A&A...573A..25S}, \citet{2015A&A...573A..26S}, \citet{2015A&A...573A..27G}, and \citet{2001ApJ...556..452L} to determine recommended present-day solar photospheric abundances. We singled out these lines to determine chemical abundances, if possible (Columns 6, 7, and 8 in Table~\ref{tab:solar}).

Photospheric solar abundances depend on a number of factors: (i) the observed spectrum, its range, resolving power, and signal-to-noise; (ii) methods of modeling the stellar atmosphere and synthetic spectra calculations, including all assumptions taken into account; (iii) the availability and quality of atomic data; (iv) the analysis method chosen to derive abundances from spectral lines. Usually two methods, equivalent width and spectrum synthesis are used, depending on the investigated line.

In the last papers considering solar chemical abundances, 3D hydrodynamical models are used, and for some elements non-LTE effects are taken into account. All methods are thoroughly described in \citet{2015A&A...573A..25S}, \citet{2015A&A...573A..26S}, and \citet{2015A&A...573A..27G} for elements from Na to Th. The investigation of abundances of all other elements is summed up by \citet{2009ARA&A..47..481A}. The solar abundances presented here were calculated for plane-parallel, LTE, 1-dimensional atmospheric models, in hydrostatic and radiative equilibrium \citep{1993KurCD..13.....K}.
Only the spectrum synthesis method was used. The atomic data, including oscillator strengths, were different for many lines analysed here and in the papers mentioned before.

\begin{figure*}
\centering
\includegraphics[width=7in]{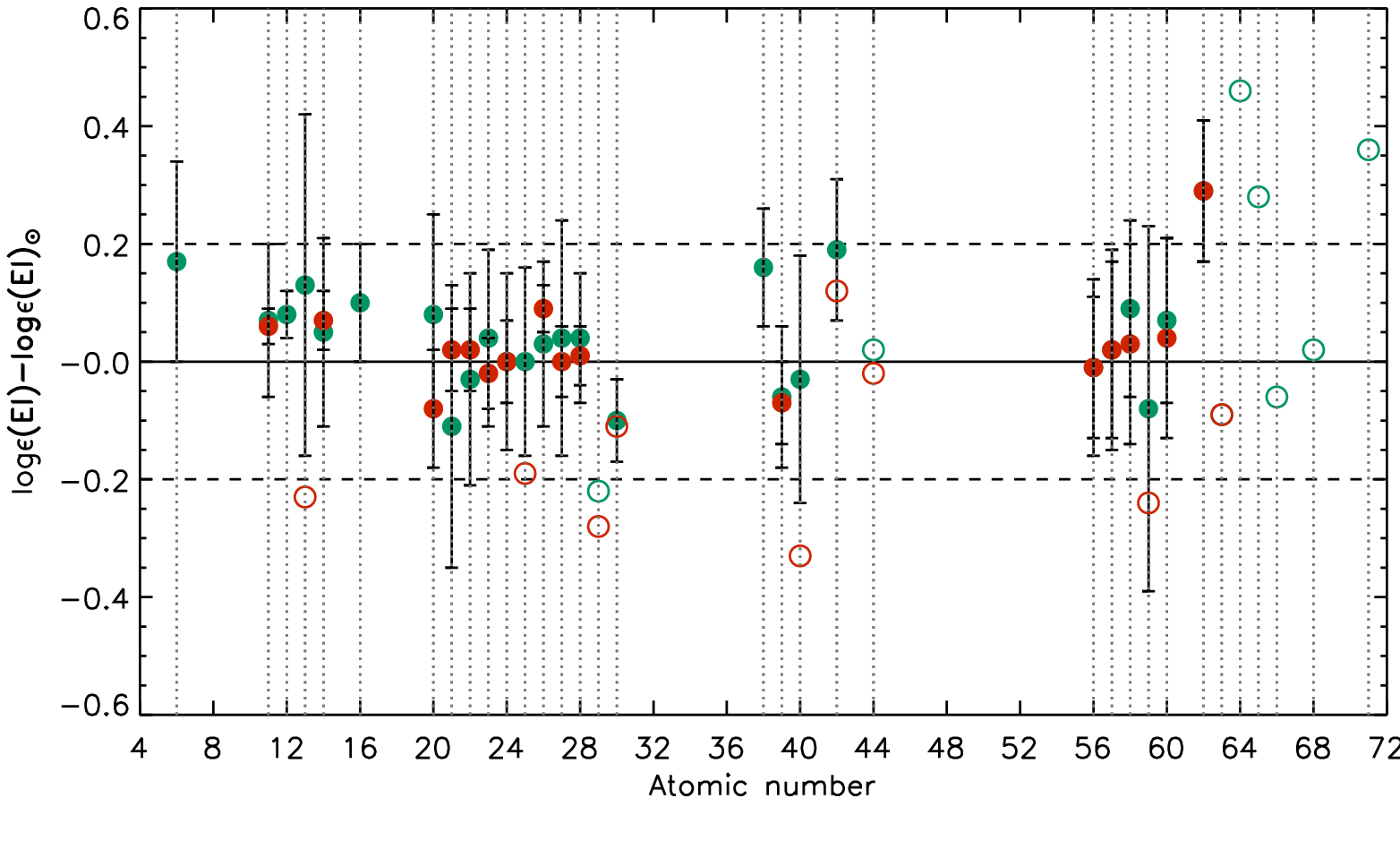}
\caption{Solar chemical abundances calculated here compared to the recommended solar values (for references see Table~\ref{tab:solar}).
Abundances determined from all lines found in the analysed spectrum are shown with green symbols, whereas
those determined from the selected lines are shown with red symbols.  
The values obtained from more than two lines are shown as filled circles, whereas those from only one or two lines are denoted as open circles.}
\label{abundances-sun}
\end{figure*}

As we can see in Fig.~\ref{abundances-sun}, the abundances of most elements agree within the error bars with the literature values. The best consistency was achieved for iron peak elements (Sc-Zn) and for the elements Na, Mg, Si, and Ca. Also for the heavy elements Y, Zr, and Mo, as well as for the rare-earth elements (REE) Ba--Nd, the obtained values are similar to the literature values. Least consistent are the results for C, O, K, and the REE Sm--Yb. Here, the carbon abundance was determined from one \ion{C}{i} line.  The value recommended by \citet{2009ARA&A..47..481A}, $\log\epsilon{\rm C} = 8.43 \pm 0.05$ derives from the mean of the 3D-based results for the [\ion{C}{I}], \ion{C}{I}, CH, and C$_2$ lines. Similarly, for the REE elements Eu--Lu only $1$ or $2$ lines were used here, instead of sets of lines analysed by \citet{2001ApJ...556..452L} for La, \citet{2009ApJS..182...51L} for Ce, \citet{2009ApJS..182...80S} for Pr, Dy, and Yb, \citet{2003ApJS..148..543D} for Nd, \citet{2006ApJS..167..292D} for Gd, \citet{2008ApJS..178...71L} for Er, and \citet{2009LanB...4B...44L} for Lu. The large differences obtained for abundances of these elements arise from a variety of sources, including a reliance on single lines of these elements, using strong lines, which increases the dependence on the adopted microturbulent velocity, or using blends where lines are contaminated by other elements.

\begin{figure*}
\centering
\includegraphics[width=7in]{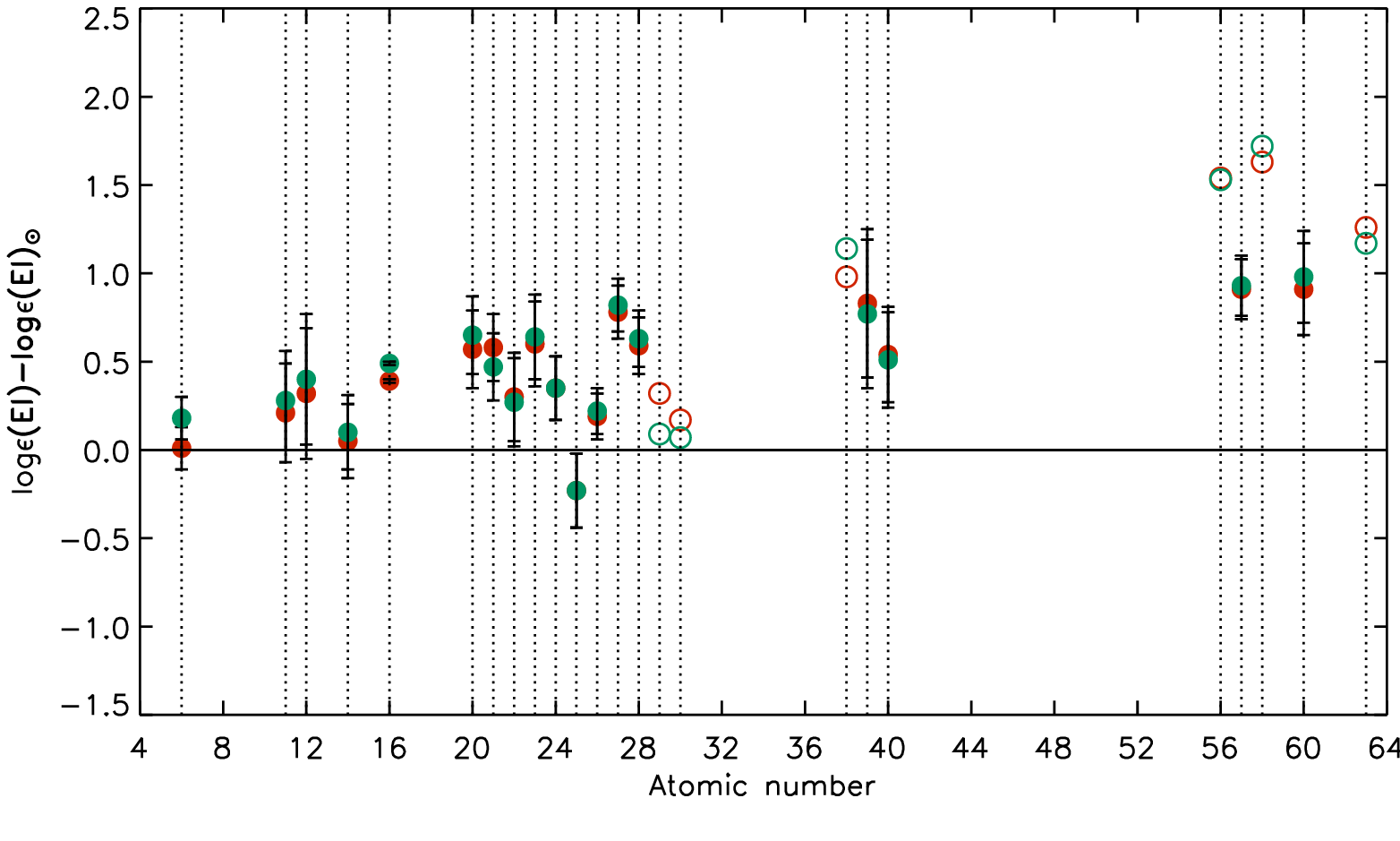}
\caption[]{Chemical abundances of HD\,21190 compared to the recommended solar values (green symbols, see Table~\ref{tab:abundances})
and solar abundances obtained here from all lines available in the analysed solar spectrum (red symbols). The values 
determined from more than two lines are shown as filled circles, whereas those from only one or two lines are denoted as open circles.
The solid line shows solar metallicity. }
\label{abundances-sun-f}
\end{figure*}

\begin{figure*}
\centering
\includegraphics[width=7in]{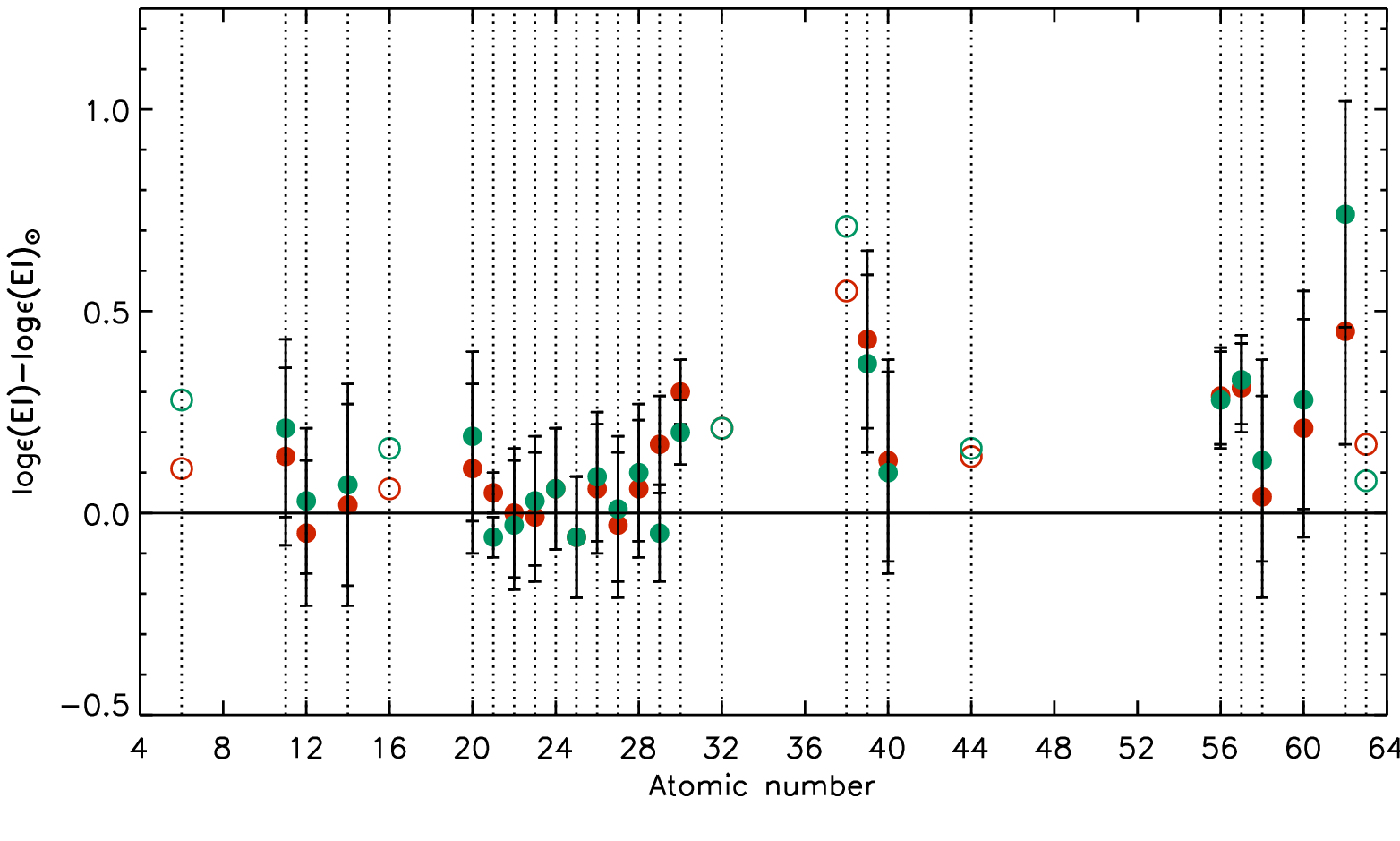}
\caption[]{Chemical abundances of CPD\,$-$83$^{\circ}$\,64B compared to the recommended solar values (green symbols, see Table~\ref{tab:abundances})
and solar abundances obtained here from all lines available in the analysed solar spectrum (red symbols).
The values determined from more than two lines are shown as filled circles, whereas those from only one or two lines are denoted as open circles.
The solid line shows solar metallicity. }
\label{abundances-sun-g}
\end{figure*}

Summarizing, we have to be careful comparing abundances of stellar atmospheres with solar values. This is mainly the case for rare-earth elements for which typically only a few weak blends are available for analysis. On the other hand, the solar abundances of light and iron-peak elements are well reproduced even for lines not considered in the papers mentioned before. The comparison of HD\,21190 and CPD\,$-$83$^{\circ}$\,64B abundances with literature solar values and solar abundances obtained in this work are shown in Figs.~\ref{abundances-sun-f} and \ref{abundances-sun-g}, respectively. In summary, the abundance patterns are similar between those obtained for HD\,21190 and those for CPD\,$-$83$^{\circ}$\,64B, no matter which solar abundances we use. 


\begin{table*}
\centering
\caption{Solar abundances of chemical elements. For chemical elements listed in the first two columns, abundances determined from
all lines found in the analysed spectrum  are shown in Columns~3, 4, and 5.
Abundances determined from the selected lines used to determine the recommended present-day solar photospheric abundances are presented in Columns~6, 7, and 8.
An average value is given; the standard deviations were calculated only if the number of the analysed lines is greater than two.
The number of lines analysed is also indicated.}
\label{tab:solar}
\begin{tabular}{lrcrcccrcccc}
\toprule
El. &Atomic & \vline & No. of & Average   & Standard  & \vline & No. of & Average   & Standard  & \vline & Solar  \\
    &number & \vline & lines  & abundance & deviation & \vline & lines  & abundance & deviation & \vline & abundances$^*$ \\
\midrule
 & & \vline & \multicolumn{3}{c}{Sun (all lines)}&  \vline & \multicolumn{3}{c}{Sun (selected lines)} & \vline \\
\midrule
 C  &  6  & \vline & $ 3 $& $ 8.60 $& $ 0.17 $ & \vline &        &         &         & \vline & $8.43\pm0.05$  \\
 Na & 11  & \vline & $ 6 $& $ 6.28 $& $ 0.13 $ & \vline &$   3  $&$  6.27 $&$  0.03$ & \vline & $6.21\pm0.04$  \\
 Mg & 12  & \vline & $ 3 $& $ 7.67 $& $ 0.04 $ & \vline &$      $&$       $&$      $ & \vline & $7.59\pm0.04$  \\
 Al & 13  & \vline & $ 6 $& $ 6.56 $& $ 0.29 $ & \vline &$   2  $&$  6.20 $&$  -   $ & \vline & $6.43\pm0.04$  \\
 Si & 14  & \vline & $36 $& $ 7.56 $& $ 0.16 $ & \vline &$   5  $&$  7.58 $&$  0.05$ & \vline & $7.51\pm0.03$  \\
 S  & 16  & \vline & $ 3 $& $ 7.22 $& $ 0.10 $ & \vline &$      $&$       $&$      $ & \vline & $7.12\pm0.03$  \\
 Ca & 20  & \vline & $36 $& $ 6.40 $& $ 0.17 $ & \vline &$   5  $&$  6.24 $&$  0.10$ & \vline & $6.32\pm0.03$  \\
 Sc & 21  & \vline & $19 $& $ 3.05 $& $ 0.24 $ & \vline &$   4  $&$  3.18 $&$  0.07$ & \vline & $3.16\pm0.04$  \\
 Ti & 22  & \vline & $167$& $ 4.90 $& $ 0.18 $ & \vline &$   9  $&$  4.95 $&$  0.07$ & \vline & $4.93\pm0.04$  \\
 V  & 23  & \vline & $59 $& $ 3.93 $& $ 0.15 $ & \vline &$  11  $&$  3.87 $&$  0.06$ & \vline & $3.89\pm0.08$  \\
 Cr & 24  & \vline & $135$& $ 5.62 $& $ 0.15 $ & \vline &$  12  $&$  5.62 $&$  0.07$ & \vline & $5.62\pm0.04$  \\
 Mn & 25  & \vline & $44 $& $ 5.42 $& $ 0.16 $ & \vline &$   2  $&$  5.23 $&$  -   $ & \vline & $5.42\pm0.04$  \\
 Fe & 26  & \vline & $524$& $ 7.50 $& $ 0.14 $ & \vline &$   3  $&$  7.56 $&$  0.04$ & \vline & $7.47\pm0.04$  \\
 Co & 27  & \vline & $59 $& $ 4.97 $& $ 0.20 $ & \vline &$   5  $&$  4.93 $&$  0.06$ & \vline & $4.93\pm0.05$  \\
 Ni & 28  & \vline & $132$& $ 6.24 $& $ 0.11 $ & \vline &$   3  $&$  6.21 $&$  0.05$ & \vline & $6.20\pm0.04$  \\
 Cu & 29  & \vline & $ 2 $& $ 3.96 $& $ -    $ & \vline &$   1  $&$  3.90 $&$  -   $ & \vline & $4.18\pm0.05$  \\
 Zn & 30  & \vline & $ 3 $& $ 4.46 $& $ 0.07 $ & \vline &$   2  $&$  4.45 $&$  -   $ & \vline & $4.56\pm0.05$  \\
 Sr & 38  & \vline & $ 2 $& $ 2.99 $& $ 0.10 $ & \vline &$      $&$       $&$      $ & \vline & $2.83\pm0.06$  \\
 Y  & 39  & \vline & $10 $& $ 2.15 $& $ 0.12 $ & \vline &$   4  $&$  2.14 $&$  0.07$ & \vline & $2.21\pm0.05$  \\
 Zr & 40  & \vline & $10 $& $ 2.56 $& $ 0.21 $ & \vline &$   1  $&$  2.26 $&$  -   $ & \vline & $2.59\pm0.04$  \\
 Mo & 42  & \vline & $ 3 $& $ 2.07 $& $ 0.12 $ & \vline &$   2  $&$  2.00 $&$  -   $ & \vline & $1.88\pm0.08$  \\
 Ru & 44  & \vline & $ 2 $& $ 1.77 $& $ -    $ & \vline &$   1  $&$  1.73 $&$  -   $ & \vline & $1.75\pm0.08$  \\
 Ba & 56  & \vline & $ 4 $& $ 2.24 $& $ 0.12 $ & \vline &$   3  $&$  2.24 $&$  0.15$ & \vline & $2.25\pm0.07$  \\
 La & 57  & \vline & $13 $& $ 1.13 $& $ 0.17 $ & \vline &$   3  $&$  1.13 $&$  0.15$ & \vline & $1.11\pm0.04$  \\
 Ce & 58  & \vline & $17 $& $ 1.67 $& $ 0.15 $ & \vline &$   7  $&$  1.61 $&$  0.17$ & \vline & $1.58\pm0.04$  \\
 Pr & 59  & \vline & $ 4 $& $ 0.64 $& $ 0.31 $ & \vline &$   1  $&$  0.48 $&$  -   $ & \vline & $0.72\pm0.04$  \\
 Nd & 60  & \vline & $21 $& $ 1.49 $& $ 0.14 $ & \vline &$   5  $&$  1.46 $&$  0.17$ & \vline & $1.42\pm0.04$  \\
 Sm & 62  & \vline & $ 7 $& $ 1.24 $& $ 0.12 $ & \vline &$   3  $&$  1.24 $&$  0.12$ & \vline & $0.95\pm0.04$  \\
 Eu & 63  & \vline & $ 1 $& $ 0.43 $& $ -    $ & \vline &$   1  $&$  0.43 $&$  -   $ & \vline & $0.52\pm0.04$  \\
 Gd & 64  & \vline & $ 2 $& $ 1.54 $& $ -    $ & \vline &$      $&$       $&$      $ & \vline & $1.08\pm0.04$  \\
 Tb & 65  & \vline & $ 1 $& $ 0.59 $& $ -    $ & \vline &$      $&$       $&$      $ & \vline & $0.31\pm0.10$  \\
 Dy & 66  & \vline & $ 2 $& $ 1.04 $& $ -    $ & \vline &$      $&$       $&$      $ & \vline & $1.10\pm0.04$  \\
 Er & 68  & \vline & $ 2 $& $ 0.95 $& $ -    $ & \vline &$      $&$       $&$      $ & \vline & $0.93\pm0.05$  \\
 Lu & 71  & \vline & $ 1 $& $ 0.46 $& $ -    $ & \vline &$      $&$       $&$      $ & \vline & $0.10\pm0.09$  \\
\bottomrule
\end{tabular}
\begin{minipage}{\textwidth}
{\bf Notes:}\\
$^*$ Solar abundances:
C -- \citet{2009ARA&A..47..481A};
Na--Ca -- \citet{2015A&A...573A..25S};
Sc--Ni -- \citet{2015A&A...573A..26S};
Cu--Er -- \citet{2015A&A...573A..27G};
Lu -- \citet{2009LanB...4B...44L}.
\end{minipage}
\end{table*}

\label{lastpage}


\begin{thebibliography}{99}

\bibitem[\protect\citeauthoryear{Andrews, Chanam{\'e}, \& Ag{\"u}eros}{2017}]{2017arXiv170407829A} Andrews J.~J., Chanam{\'e} J., Ag{\"u}eros M.~A., 2017, arXiv, arXiv:1704.07829

\bibitem[\protect\citeauthoryear{Appenzeller et al.}{1998}]{Appenzeller1998} Appenzeller I., et al., 1998, The Messenger, 94, 1

\bibitem[\protect\citeauthoryear{Asplund et al.}{2009}]{2009ARA&A..47..481A} Asplund M., Grevesse N., Sauval A.~J., Scott P., 2009, ARA\&A, 47, 481 

\bibitem[\protect\citeauthoryear{Balona et al.}{2011}]{2011MNRAS.410..517B}	Balona L.~A., et al., 2011, MNRAS, 410, 517

\bibitem[\protect\citeauthoryear{Bland-Hawthorn \& Gerhard}{2016}]{BlandHawthorn16} Bland-Hawthorn J., Gerhard O., 2016, ARA\&A, 54, 529

\bibitem[\protect\citeauthoryear{Caliskan}{2005}]{2005MNRAS.356.1256C} Caliskan H., 2005, MNRAS, 356, 1256

\bibitem[\protect\citeauthoryear{Casagrande et al.}{2011}]{2011A&A...530A.138C} Casagrande L., Sch{\"o}nrich R., Asplund M., Cassisi S., Ram{\'{\i}}rez I., Mel{\'e}ndez J., Bensby T., Feltzing S., 2011, A\&A, 530, A138 

\bibitem[\protect\citeauthoryear{Catanzaro, Leone, \& Dall}{2004}]{2004A&A...425..641C} Catanzaro G., Leone F., Dall T.~H., 2004, A\&A, 425, 641 

\bibitem[\protect\citeauthoryear{Cutri et al.}{2003}]{2003yCat.2246....0C} Cutri R.~M., et al., 2003, ``2MASS All-Sky Catalog of Point Sources'', VizieR Online Data Catalog 2246

\bibitem[\protect\citeauthoryear{Dall et al.}{2006}]{2006A&A...454..341D} Dall T.~H., Santos N.~C., Arentoft T., Bedding T.~R., Kjeldsen H., 2006, A\&A, 454, 341 

\bibitem[\protect\citeauthoryear{Den Hartog et al.}{2003}]{2003ApJS..148..543D} Den Hartog E.~A., Lawler J.~E., Sneden C., Cowan J.~J., 2003, ApJS, 148, 543 

\bibitem[\protect\citeauthoryear{Den Hartog et al.}{2006}]{2006ApJS..167..292D} Den Hartog E.~A., Lawler J.~E., Sneden C., Cowan J.~J., 2006, ApJS, 167, 292

\bibitem[\protect\citeauthoryear{Dommanget \& Nys}{2000}]{dommanget00} Dommanget J., Nys O., 2000, A\&A, 363, 991 

\bibitem[\protect\citeauthoryear{Doyle et al.}{2014}]{2014MNRAS.444.3592D} Doyle A.~P., Davies G.~R., Smalley B., Chaplin W.~J., Elsworth Y., 2014, MNRAS, 444, 3592

\bibitem[\protect\citeauthoryear{ESA}{1997}]{esa97} ESA 1997, The HIPPARCOS and Tycho catalogues, ESA-SP~1200

\bibitem[\protect\citeauthoryear{Gaia Collaboration et al.}{2016a}]{gaia16a} Gaia Collaboration, et al., 2016a, A\&A, 595, A1 

\bibitem[\protect\citeauthoryear{Gaia Collaboration et al.}{2016b}]{gaia16b} Gaia Collaboration, et al., 2016b, A\&A, 595, A2

\bibitem[\protect\citeauthoryear{Girard et al.}{2011}]{girard11} Girard T.~M., 2011, AJ, 142, 15

\bibitem[\protect\citeauthoryear{Gonz{\'a}lez, Hubrig, \& Savanov}{2008}]{2008CoSka..38..411G} Gonz{\'a}lez J.~F., Hubrig S., Savanov I., 2008, CoSka, 38, 411 
\bibitem[\protect\citeauthoryear{Gonz{\'a}lez et al.}{2008}]{2008MNRAS.384.1140G} Gonz{\'a}lez J.~F., Hubrig S., Kurtz D.~W., Elkin V., Savanov I., 2008, MNRAS, 384, 1140

\bibitem[\protect\citeauthoryear{Gray}{2005}]{gray} Gray D.~F., 2005, {\it The Observation and Analysis of Stellar Photospheres}, Cambridge University Press

\bibitem[\protect\citeauthoryear{Gray \& Corbally}{2009}]{2009ssc..book.....G} Gray R.~O., Corbally J.~C., 2009, Stellar Spectral Classification by Richard O.~Gray and Christopher J.~Corbally.~Princeton University Press

\bibitem[\protect\citeauthoryear{Gray, Graham, \& Hoyt}{2001}]{2001AJ....121.2159G} Gray R.~O., Graham P.~W., Hoyt S.~R., 2001, AJ, 121, 2159  

\bibitem[\protect\citeauthoryear{Grevesse et al.}{2015}]{2015A&A...573A..27G} Grevesse N., Scott P., Asplund M., Sauval A.~J., 2015, A\&A, 573, A27

\bibitem[\protect\citeauthoryear{Grigahc{\`e}ne et al.}{2005}]{2005A&A...434.1055G} Grigahc{\`e}ne A., Dupret M.-A., Gabriel M., Garrido R., Scuflaire R., 2005, A\&A, 434, 1055 

\bibitem[\protect\citeauthoryear{Hartkopf et al.}{2013}]{hartkopf13} Hartkopf W.~I., Mason B.~D., Finch C.~T., Zacharias N., Wycoff G.~L., Hsu D., 2013, AJ, 146, 76

\bibitem[\protect\citeauthoryear{Hauck \& Mermilliod}{1998}]{1998A&AS..129..431H} Hauck B., Mermilliod M., 1998, A\&AS, 129, 431 

\bibitem[\protect\citeauthoryear{H{\o}g et al.}{1998}]{hog98} H{\o}g E., Kuzmin A., Bastian U., Fabricius C., Kuimov K., Lindegren L., Makarov V.~V., Roeser S., 1998, A\&A, 335, L65

\bibitem[\protect\citeauthoryear{H{\o}g et al.}{2000}]{2000A&A...355L..27H} H{\o}g E., et al., 2000, A\&A, 355, L27 

\bibitem[\protect\citeauthoryear{Hubrig, North, \& Mathys}{2000}]{2000ApJ...539..352H} Hubrig S., North P., Mathys G., 2000, ApJ, 539, 352

\bibitem[\protect\citeauthoryear{Hubrig \& Sch\"oller}{2016}]{2016IBVS.6174....1H} Hubrig S., Sch\"oller M., 2016, IBVS, 6174, 1 

\bibitem[\protect\citeauthoryear{Johnson \& Soderblom}{1987}]{johnson87} Johnson D.~R. H., Soderblom D.~R., 1987, AJ, 93, 864

\bibitem[\protect\citeauthoryear{Kharchenko}{2001}]{kharchenko01} Kharchenko N.~V., 2001, Kinematika i Fizika Nebesnykh Tel, 17, 409 

\bibitem[\protect\citeauthoryear{Koen et al.}{2001}]{2001MNRAS.326..387K} Koen C., Kurtz D.~W., Gray R.~O., Kilkenny D., Handler G., Van Wyk F., Marang F., Winkler H., 2001, MNRAS, 326, 387 

\bibitem[\protect\citeauthoryear{Kunder et al.}{2017}]{2017AJ....153...75K} Kunder A., et al., 2017, AJ, 153, 75

\bibitem[\protect\citeauthoryear{Kurtz et al.}{2006}]{2006MNRAS.370.1274K} Kurtz D.~W., Elkin V.~G., Mathys G., 2006, MNRAS, 370, 1274

\bibitem[\protect\citeauthoryear{Kurtz et al.}{2008}]{2008MNRAS.386.1750K} Kurtz D.~W., Hubrig S., Gonzalez J.~F., van Wyk F., Martinez P., 2008, MNRAS, 386, 1750

\bibitem[\protect\citeauthoryear{Kurucz}{1993}]{1993KurCD..13.....K} Kurucz R., 1993, ATLAS9 Stellar Atmosphere Programs and 2~km/s grid. Kurucz CD-ROM No.~13. Cambridge, Mass.: Smithsonian Astrophysical Observatory, 13,

\bibitem[\protect\citeauthoryear{Kurucz}{1993b}]{1993KurCD..18.....K} Kurucz R.~L., 1993, Kurucz CD-ROM 18, SAO, Cambridge, USA.

\bibitem[\protect\citeauthoryear{Kurucz}{2014}]{2014dapb.book...63K} Kurucz R.~L., 2014, in {\it Determination of Atmospheric Parameters of B-, A-, F- and G-Type Stars.}, Series: GeoPlanet: Earth and Planetary Sciences, ISBN: 978-3-319-06955-5. Springer International Publishing, Edited by Ewa Niemczura, Barry Smalley and Wojtek Pych, pp. 63-73 

\bibitem[\protect\citeauthoryear{Lawler, Bonvallet, \& Sneden}{2001}]{2001ApJ...556..452L} Lawler J.~E., Bonvallet G., Sneden C., 2001, ApJ, 556, 452

\bibitem[\protect\citeauthoryear{Lawler et al.}{2008}]{2008ApJS..178...71L} Lawler J.~E., Sneden C., Cowan J.~J., Wyart J.-F., Ivans I.~I., Sobeck J.~S., Stockett M.~H., Den Hartog E.~A., 2008, ApJS, 178, 71-88

\bibitem[\protect\citeauthoryear{Lawler et al.}{2009}]{2009ApJS..182...51L} Lawler J.~E., Sneden C., Cowan J.~J., Ivans I.~I., Den Hartog E.~A., 2009, ApJS, 182, 51

\bibitem[\protect\citeauthoryear{Liakos \& Niarchos}{2017}]{liakos16} Liakos A., Niarchos P., 2017, MNRAS, 465, 1181

\bibitem[\protect\citeauthoryear{Lindegren et al.}{2016}]{lindegren16} Lindegren L., et al., 2016, A\&A, 595, A4

\bibitem[\protect\citeauthoryear{Lodders, Palme, \& Gail}{2009}]{2009LanB...4B...44L} Lodders K., Palme H., Gail H.-P., 2009, Landolt-B\"ornstein -- Group VI, Springer-Verlag, Berlin/Heidelberg

\bibitem[\protect\citeauthoryear{L{\"u}}{1971}]{lu71} L{\"u} P.~K., 1971, Transactions of the Astronomical Observatory of Yale University, 31, 1

\bibitem[\protect\citeauthoryear{Masana, Jordi, \& Ribas}{2006}]{2006A&A...450..735M} Masana E., Jordi C., Ribas I., 2006, A\&A, 450, 735 

\bibitem[\protect\citeauthoryear{Mason et al.}{2001}]{mason01} Mason B.~D., Wycoff G.~L., Hartkopf W.~I., Douglass G.~G., Worley C.~E., 2001, AJ, 122, 3466

\bibitem[\protect\citeauthoryear{Mayor et al.}{2003}]{mayor2003} Mayor M. et al. 2003, The Messenger 114, 20

\bibitem[\protect\citeauthoryear{McDonald, Zijlstra, \& Boyer}{2012}]{2012MNRAS.427..343M} McDonald I., Zijlstra A.~A., Boyer M.~L., 2012, MNRAS, 427, 343 

\bibitem[\protect\citeauthoryear{Moon \& Dworetsky}{1985}]{1985MNRAS.217..305M} Moon T.~T., Dworetsky M.~M., 1985, MNRAS, 217, 305 

\bibitem[\protect\citeauthoryear{Munari \& Zwitter}{1997}]{1997A&A...318..269M} Munari U., Zwitter T., 1997, A\&A, 318, 269

\bibitem[\protect\citeauthoryear{Napiwotzki, Schoenberner, \& Wenske}{1993}]{1993A&A...268..653N} Napiwotzki R., Schoenberner D., Wenske V., 1993, A\&A, 268, 653 

\bibitem[\protect\citeauthoryear{Neiner \& Lampens}{2015}]{2015MNRAS.454L..86N} Neiner, C., Lampens, P., 2015, MNRAS, 454, L86

\bibitem[\protect\citeauthoryear{Niemczura et al.}{2015}]{2015MNRAS.450.2764N} Niemczura E., et al., 2015, MNRAS, 450, 2764

\bibitem[\protect\citeauthoryear{Pamyatnykh}{2000}]{2000ASPC..210..215P} Pamyatnykh A.~A., 2000, ASPC, 210, 215 

\bibitem[\protect\citeauthoryear{Rousseau, Perie, \& Gachard}{1996}]{rousseau96} Rousseau J.~M., Perie J.~P., Gachard M.~T., 1996, A\&A Suppl.\ Ser., 116, 301
 
\bibitem[\protect\citeauthoryear{Saio}{2005}]{2005MNRAS.360.1022S} Saio H., 2005, MNRAS, 360, 1022S 

\bibitem[\protect\citeauthoryear{Savage \& Mathis}{1979}]{1979ARA&A..17...73S} Savage B.~D., Mathis J.~S., 1979, ARA\&A, 17, 73

\bibitem[\protect\citeauthoryear{Sbordone}{2005}]{sbordone} Sbordone L., 2005, Memorie della Societ\`a Astronomica Italiana Suppl., 8, 61

\bibitem[\protect\citeauthoryear{Scott et al.}{2015a}]{2015A&A...573A..25S} Scott P., et al., 2015a, A\&A, 573, A25 

\bibitem[\protect\citeauthoryear{Scott et al.}{2015b}]{2015A&A...573A..26S} Scott P., Asplund M., Grevesse N., Bergemann M., Sauval A.~J., 2015b, A\&A, 573, A26 

\bibitem[\protect\citeauthoryear{Skrutskie et al.}{2006}]{skrutskie06} Skrutskie M.~F., et al., 2006, AJ, 131, 1163

\bibitem[\protect\citeauthoryear{Smalley}{2014}]{2014dapb.book...85S} Smalley B., 2014, dapb.book, 85 

\bibitem[\protect\citeauthoryear{Sneden et al.}{2009}]{2009ApJS..182...80S} Sneden C., Lawler J.~E., Cowan J.~J., Ivans I.~I., Den Hartog E.~A., 2009, ApJS, 182, 80

\bibitem[\protect\citeauthoryear{Starikova}{1979}]{starikova79} Starikova G.~A., 1979, Soviet Astronomy Letters, 5, 353

\bibitem[\protect\citeauthoryear{Urban et al.}{1998a}]{urban98a} Urban S.~E., Corbin T.~E., Wycoff G.~L., Martin J.~C., Jackson E.~S., Zacharias M.~I., Hall D.~M., 1998a, AJ, 115, 1212

\bibitem[\protect\citeauthoryear{Urban, Corbin, \& Wycoff}{1998b}]{urban98b} Urban S.~E., Corbin T.~E., Wycoff G.~L., 1998b, AJ, 115, 2161

\bibitem[\protect\citeauthoryear{van Leeuwen}{2007}]{vanleeuwen07} van Leeuwen F., 2007, A\&A, 474, 653

\bibitem[\protect\citeauthoryear{Wright}{2010}]{wright10} Wright E.~L., et al., 2010, AJ, 140, 1868

\bibitem[\protect\citeauthoryear{Wycoff, Mason, \& Urban}{2006}]{wycoff06} Wycoff G.~L., Mason B.~D., Urban S.~E., 2006, AJ, 132, 50

\bibitem[\protect\citeauthoryear{Zacharias et al.}{2013}]{zacharias13} Zacharias N., Finch C.~T., Girard T.~M., Henden A., Bartlett J.~L., Monet D.~G., Zacharias M.~I., 2013, AJ, 145, 44

\bibitem[\protect\citeauthoryear{Zacharias, Finch, \& Frouard}{2017}]{2017arXiv170205621Z} Zacharias, N., Finch, C., \& Frouard, J.\ 2017, AJ, 153, 166 

\end{thebibliography}
\end{document}